\documentclass[10pt,twocolumn,english]{revtex4-2}
\usepackage[T1]{fontenc}
\usepackage[latin9]{inputenc}
\usepackage[active]{srcltx}
\usepackage{babel}
\usepackage{amstext}
\usepackage{amssymb}
\usepackage{graphicx}
\usepackage{wasysym}
\usepackage[unicode=true,pdfusetitle,
 bookmarks=true,bookmarksnumbered=false,bookmarksopen=false,
 breaklinks=false,pdfborder={0 0 1},backref=false,colorlinks=false]
 {hyperref}

\makeatletter

\providecommand{\tabularnewline}{\\}

\makeatother

\begin{document}
\title{Non-equilibrium Ising Model on a 2D \textit{Additive} Small-World
Network}
\author{R. A. Dumer}
\affiliation{Instituto de Física - Universidade Federal de Mato Grosso, 78060-900,
Cuiabá MT, Brazil.}
\author{M. Godoy}
\affiliation{Instituto de Física - Universidade Federal de Mato Grosso, 78060-900,
Cuiabá MT, Brazil.}
\begin{abstract}
In this work, we have studied the Ising model with one- and two-spin
flip competing dynamics on a two-dimensional \emph{additive} small-world
network (A-SWN). The system model consists of a $L\times L$ square
lattice where each site of the lattice is occupied by a spin variable
that interacts with the nearest neighbor spins and it has a certain
probability $p$ of being additionally connected at random to one
of its farther neighbors. The dynamics present in the system can be
defined by the probability $q$ of being in contact with a heat bath
at a given temperature $T$ and, at the same time, with a probability
of $1-q$ the system is subjected to an external flux of energy into
the system. The contact with the heat bath is simulated by one-spin
flip according to the Metropolis prescription, while the input of
energy is mimicked by the two-spin flip process, involving a simultaneous
flipping of a pair of neighboring spins. We have employed Monte Carlo
simulations to obtain the thermodynamic quantities of the system,
such as, the total $\textrm{\ensuremath{\textrm{m}_{\textrm{L}}^{\textrm{F}}}}$
and staggered $\textrm{\ensuremath{\textrm{m}_{\textrm{L}}^{\textrm{AF}}}}$
magnetizations per spin, the susceptibility $\textrm{\ensuremath{\chi_{\textrm{L}}}}$,
and the reduced fourth-order Binder cumulant $\textrm{\ensuremath{\textrm{U}_{\textrm{L}}}}$.
We have built the phase diagram for the stationary states of the model
in the plane $T$ versus $q$, showing the existence of two continuous
transition lines for each value of $p$: one line between the ferromagnetic
$F$ and paramagnetic $P$ phases, and the other line between the
$P$ and antiferromagnetic $AF$ phases. Therefore, we have shown
that the phase diagram topology changes when $p$ increases. Using
the finite-size scaling analysis, we also obtained the critical exponents
for the system, where varying the parameter $p$, we have observed
a different universality class from the Ising model in the regular
square lattice to the A-SWN.
\end{abstract}
\keywords{Competing dynamics; Small-world network; Stationary state; Phase transitions;}
\maketitle

\subsection{Introduction}

\begin{figure*}
\begin{centering}
\includegraphics{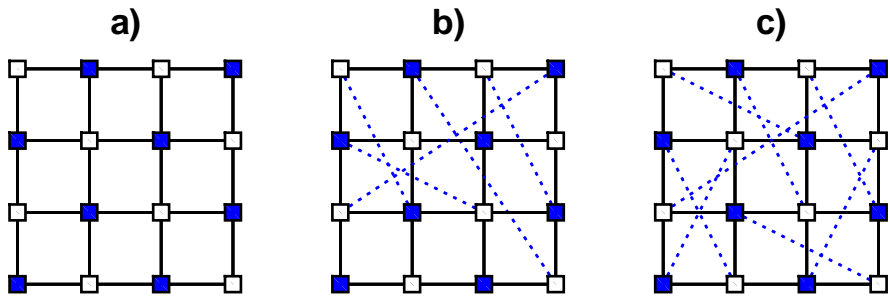}
\par\end{centering}
\caption{{\footnotesize{}Schematic representation of the system and the A-SWN.
The blue square dots indicate the sites on one of the sublattices,
the white square dots are the sites on the other sublattice, the solid
black lines are the nearest-neighbor interactions $J$ between pairs
of spins, and the blue-dashed lines are long-range interaction $J_{ik}$
added to the network with a certain probability $p$. (a) for $p=0$,
(b) for $p=0.5$ and (c) for $p=1$.\label{fig:1}}}
\end{figure*}

In the 1960s, the dynamic behavior of the Ising model was successfully
described by Glauber \citep{1} and Kawasaki \citep{2} mechanisms.
This instigated interest in the competition between the Glauber and
Kawasaki stochastic process, one-spin flip and two-spin exchange,
respectively, in this model. This competition can be simulated by
the Glauber process with probability $q$ simulating the system in
contact with a heat bath at a temperature $T$, and at the same time,
with probability $1-q$, the Kawasaki process mimics an input of energy
into the system \citep{3}. Each of these dynamical processes singly
satisfies the detailed balance condition, which drives the system
toward equilibrium. However, when both act simultaneously, the detailed
balance is no longer satisfied and the system is forced out of equilibrium.

The Ising model on a regular square lattice has the critical temperature
and universality class are given by the critical exponents well-known
exactly at the equilibrium state \citep{4}. Therefore, the stationary
non-equilibrium states were obtained by the two competing dynamic
processes described above, and a self-organization is observed by
the disappearance of the ordered ferromagnetic $F$ phase in the transition
to the paramagnetic $P$ phase, and identification of the ordered
antiferromagnetic $AF$ phase, as we increase the flow of energy into
the system \citep{5}. However, through the Monte Carlo simulations
(MC) the critical exponents of the system have been obtained, and
because it is a system with the same symmetry, spatial dimension,
and range interactions, the exponents are the same as at the equilibrium
state model, and known exactly \citep{6}. In the same way, Godoy
and Figueredo investigated the mixed-spin Ising model, which does
not admit spin exchanges between the spin sublattices, consequently
do not admit to utilizing the Kawasaki dynamic. Thus, the competing
dynamic was made by the one- and two-spin flip mechanisms, and even
with that, they have also obtained the self-organization phenomena
\citep{7}, and based on the critical behavior of the system, the
universality class of the system is the same that the Ising model
with only spin-1/2 \citep{8}. Therefore, in the non-equilibrium models,
the universality class of the stationary critical behavior is the
same as in the equilibrium models. All of these works were studied
on regular square lattices.

By using graph theory, Watts and Strogatz quantify the properties
of Small-World phenomena as demonstrated in Milgram's 1967 study \citep{9}.
As an underlying assumption of the Watts-Strogatz model (WS-model)
\citep{10}, vertices of graphs are sites of networks, and edges are
connections between sites of the networks. By introducing a disorder
parameter $p$, as the probability of randomly rewriting each one
of the connections in a regular lattice, we can obtain the SWN in
specific regions in the interval $0<p\le1$. The SWN regime is identified
in regions of $p$ where the network possesses local clustering, $C(p)$,
of a regular lattice, but at the same time has an average distance
between any two sites, $l(p)$, characteristic of a random lattice.
In addition to the WS-model, some variants of this model were also
developed to describe the properties of an SWN. One of these variants
\citep{11} uses a regular square lattice, and we can add a long-range
interaction to each site with a certain probability $p$. This leads
to a small typical separation, preserving the clustering property
of a regular lattice. While we have described the rewiring SWN (R-SWN)
in the WS-model \citep{10}, this last form is known as additive SWN
(A-SWN) \citep{11}.

These networks have been used in numerous physical models since the
initial SWN model was put forth \citep{12,13,14,15,16}, including
the Ising model in 1D, 2D, and 3D for the investigation of the critical
phenomena at equilibrium system \citep{17,18,19,20,21,22,23,24}.
According to these findings for the Ising model, an order to the disorder
phase transition is established for $T\ne0$ with $0<p\le1$, and
it is seen that the addition of long-range interactions changes the
critical behavior of the system.

The interesting behavior of the Ising model at the equilibrium SWN,
its investigation was also carried out about the non-equilibrium phase
transitions by the competing dynamics: analytically in 1D \citep{25},
by MC simulations in 2D \citep{26}, and by the Gaussian model in
3D \citep{27}. In all of these works they have been using the competition
between the Glauber and Kawasaki dynamics, and have no conclusions
about the mean-field critical behavior observed at the equilibrium
Ising model on an SWN \citep{18,20,21,22,24,key-1}. However, in 2D
and 3D systems, is obtained the $AF-P$ and $F-P$ phase transitions,
characteristic of the self-organization phenomena, and observed in
all of the other systems at the non-equilibrium state by the competing
dynamics.

In the present work, we have investigated the Ising model in a two-dimensional
A-SWN, where each site of the network is occupied by a spin variable
spin-1/2 that can assume values $\pm1$. We limit by one the number
of long-range interactions that each site can receive with probability
$p$, and divide the network into two sublattices, each new interaction
created should connect these sublattices. The system is in a non-equilibrium
regime by competing between two dynamic processes that do not conserve
the order parameter: with competition probability $q$, the one-spin
flip process simulates the system in contact with a heat bath at temperature
$T$, and with competition probability $1-q$ the two-spin flip process
mimics the system subjected to an external energy flux into it. Therefore,
the system is studied at the non-equilibrium regime due to competing
dynamics. We verified the phase transition between the $AF$ and $F$
ordered phases to the $P$ disordered phase, and if the system is
in this A-SWN regime, it exhibit the same mean-field critical behavior
observed at equilibrium systems with long-range interactions by the
A-SWN, see Ref. \citep{key-1}. The behavior of the phase transitions,
phases diagrams and critical exponents by FSS analysis also are described
and compared with those of Ref. \citep{key-1}.

This work is organized as follows: In Section \ref{subsec:Model},
we describe the model, the network, and the motion equations for the
non-equilibrium Ising model. In Section \ref{subsec:Monte-Carlo-simulations},
we present the MC simulation method used. The behavior of the phase
transitions, phase diagrams, and critical exponents by FSS analysis
is described in Section \ref{subsec:Results}. Finally, in Section
\ref{subsec:Conclusions}, we present our conclusions.

\subsection{Model\label{subsec:Model}}

The Ising model with $N=L^{2}$ spins $\sigma_{i}=\pm1$ on a regular
square lattice $L\times L$, periodic boundary conditions, and a nearest-neighbor
ferromagnetic interaction of strength $J$ has been studied in this
work (see Fig. \ref{fig:1}(a)). On the other hand, with a certain
probability $p$, we can add one long-range interaction $J_{ik}$
to each site of that regular square lattice. We divided the system
into two sublattices to add the long-range interactions $J_{ik}$,
in which one sublattice plays the role of central spins, while the
other sublattice contains the spins in which the central spins can
connect, to beyond their nearest neighbors.\textbf{ }Thus, to choose
a long-range interaction $J_{ik}$ for a site $i$, the sublattice
of $i$ will be the sublattice of the central spins, then, we choose
randomly a site $k$ from another sublattice. If the site $k$ does
not be one of its nearest neighbors already naturally coupled with
$i$, we picked a random number $0<r<1$, and if $r\le p$ (with $p$
predefined), then we couple the site $k$ to the neighbors of site
$i$, and for the site $k$ we couple the site $i$ to its neighbors.
The attempt to add a long-range interaction $J_{ik}$ is made once
to each site that does not have a long-range interaction $J_{ik}$
in the network, and as result, we have a network with an average coordination
number $z=4+p$. Therefore, we can think as an example of some situations:
i) for $p=0$, i.e., the probability of adding a long-range interaction
$J_{ik}$ to any site on the lattice is zero, therefore, we have a
regular square lattice, see Fig \ref{fig:1}(a); ii) for $p=0.5$,
we are in the A-SWN regime because in addition to the conservation
of $C(p)$, and we also have an average short path length between
network sites, through the shortcuts created by the long-range interaction
$J_{ik}$ added between the sublattices, see Fig. \ref{fig:1}(b);
finally, for $p=1$, all sites on the network have a long-range interaction
$J_{ik}$ connecting the two sublattices, and consequently, it is
the network with the shortest typical separation between the sites
on the network, see Fig. \ref{fig:1}(c).

Thus, as the regular structure in $p=0$ keeps unaltered (Fig. \ref{fig:1}(a)),
we have a high local clustering for any value of $p$, and conform
we increase $p$, the long-range interaction $J_{ik}$ is added to
the network, creating shortcuts between the sites that before in the
simple regular lattice would be more distant, consequently decreasing
the typical distance $l(p)$ of the network. The $l(p)$ scales linearly
$l(p\to0)\sim L/2$ and logarithmically $l(p\to1)\sim\ln(L^{1.77})$,
being these regimes referred to as the ``large-world'' and ``small-world''
respectively. The cross-over between these regimes occurs when the
average number of shortcuts is about one, or in the other words, we
can say in the SWN regime when $p\apprge2L^{-2}$ \citep{5}. Versed
on this, our study is based on $p\ge0.25$ values, where the A-SWN
is found and the decay of $l$ as a function of $p$ undergoes less,
i.e., having approximately the same value of $l$.

The ferromagnetic Ising spin energy is described by the Hamiltonian
of the form:

\begin{equation}
{\cal H}=-J\sum_{\left\langle i,j\right\rangle }\sigma_{i}\sigma_{j}-\sum_{\left\langle i,k\right\rangle }J_{ik}\sigma_{i}\sigma_{k},\label{eq:1}
\end{equation}
where $J$ is the nearest-neighbor ferromagnetic interaction, and
$J_{ik}$ is the long-range interaction on the A-SWN. The first sum
is over all the pair of nearest-neighbor spins on the regular square
lattice and the second sum is made over all the pairs of spins $(i,k)$
connected through long-range interaction on the A-SWN. Here, we always
are considering $J_{ik}=J=1$.

We are dealing with the non-equilibrium Ising model and in an SWN,
being the time evolution of the states of the system governed by two
competing dynamical processes: one simulating the contact of the system
with a heat bath at temperature $T$, with the one-spin flip process
and probability $q$ to occur, and at the same time but with probability
$(1-q)$ to occur, the system is subjected to an external flux of
energy into the system with the two-spin flip process, where in addition
to flipping the chosen spin, it simultaneously flips one of its randomly
chosen neighbors. 

Let us call $p(\{\sigma\},t)$ the probability of finding the system
in the state $\{\sigma\}=\{\sigma_{1},...,\sigma_{i},...,\sigma_{j},...\sigma_{N}\}$
at time $t$, the motion equation for the probability states evolve
in time according to the master equation

\begin{equation}
\frac{d}{dt}p(\{\sigma\},t)=qG+(1-q)V,\label{eq:2}
\end{equation}
where $qG$ represents the process of relaxation of the spins in contact
with a heat bath at temperature $T$, favoring the lowest energy in
the system, and $(1-q)V$ represents the process independent of the
temperature, where the energy of the system increases by one external
flow of energy into it. $G$ and $V$ are described by

\begin{equation}
\begin{array}{ccc}
G= & \sum_{i,\{\sigma'\}}\left[W(\sigma_{i}\to\sigma_{i}')p(\{\sigma\},t)+\right.\\
 & \left.-W(\sigma_{i}'\to\sigma_{i})p(\{\sigma'\},t)\right] & ,
\end{array}\label{eq:3}
\end{equation}

\begin{equation}
\begin{array}{ccc}
V= & \sum_{i,j,\{\sigma'\}}\left[W(\sigma_{i}\sigma_{j}\to\sigma_{i}'\sigma_{j}')p(\{\sigma\},t)+\right.\\
 & \left.-W(\sigma_{i}'\sigma_{j}'\to\sigma_{i}\sigma_{j})p(\{\sigma'\},t)\right] & ,
\end{array}\label{eq:4}
\end{equation}
where $\{\sigma'\}$ denotes the spin configurations after the spin
flipping, $W(\sigma_{i}\to\sigma_{i}')$ is the transition rate between
states in the one-spin flip process, and $W(\sigma_{i}\sigma_{j}\to\sigma_{i}'\sigma_{j}')$
the transition rate between the states in the two-spin flip process,
with the order parameter being conserved in none of the dynamic processes.

If $0<q<1$, we have two dynamics processes acting simultaneously,
the detailed balance is not satisfied and the system is forced out
of equilibrium. As these processes favor the states of higher and
lower energy of the system, with the competition it is possible to
find stationary states for the order parameter in the $AF$, $F$,
and $P$ phases. It is worth noting that to reach the stationary state
in the $AF$ phase was of fundamental importance to use the $J_{ik}$
between the sublattices, because of the antiparallel ordering in which
this phase is characterized.

\subsection{Monte Carlo simulations\label{subsec:Monte-Carlo-simulations}}

Let $(k,l)$ and $(k',l')$ be the coordinates of a site in our two-dimensional
SWN and one of your neighbors respectively. The periodic boundary
conditions were used in all our simulations. Starting the initial
state of the system with all spins aligned in the same direction,
a new configuration is generated by the following the Markov process:
for a given temperature $T$, competition probability $q$, and additive
probability $p$, we choose a random spin from the lattice, i.e.,
we choose a coordinate $k$ and $l$ at random. Then we generate a
random number $\xi$ between zero and one, and if $\xi\le q$ we choose
the one-spin flip process. In this process, the flipping probability
is dependent on $W(\sigma_{kl}\to\sigma_{kl}^{\prime})$, which is
given by the Metropolis prescription as follows:

\begin{equation}
W(\sigma_{kl}\to\sigma_{kl}^{\prime})=\left\{ \begin{array}{cccc}
\exp(-\Delta E_{kl}/k_{B}T) & \textrm{if} & \Delta E_{kl}>0\\
1 & \textrm{if} & \Delta E_{kl}\le0 & ,
\end{array}\right.\label{eq:5}
\end{equation}
where $\Delta E_{kl}$ is the change in the energy after flipping
the spin $\sigma_{kl}\to\sigma_{kl}^{\prime}$, $k_{B}$ is the Boltzmann
constant, and $T$ is the absolute temperature, thus, the new state
is accepted if $\Delta E_{kl}\le0$, and in the case of $\Delta E_{kl}>0$
we choose another random number $1<\xi_{1}<0$ and if $\xi_{1}\le\exp(-\Delta E_{kl}/k_{B}T)$
the new state is also accepted, but if none of the conditions are
satisfied, we do not change the state of the system. On the other
hand, if $\xi>q$, the two-spin flip process is chosen. In this case,
in addition to the spin $\sigma_{kl}$, we also randomly choose one
of its neighbors $\sigma_{k^{\prime}l^{\prime}}$, which can be either
the nearest neighbor or the farthest neighbor coming from a $J_{ik}$.
In this process, the two spins chosen are flipping simultaneously,
and for that, the two-spin flip probability is dependent on $W(\sigma_{kl}\sigma_{k^{\prime}l^{\prime}}\to\sigma_{kl}^{\prime}\sigma_{k^{\prime}l^{\prime}}^{\prime})$,
which is given by 

\begin{equation}
W(\sigma_{kl}\sigma_{k^{\prime}l^{\prime}}\to\sigma_{kl}^{\prime}\sigma_{k^{\prime}l^{\prime}}^{\prime})=\left\{ \begin{array}{c}
0\\
1
\end{array}\begin{array}{c}
\textrm{if}\\
\textrm{if}
\end{array}\begin{array}{cc}
\Delta E_{kl,k'l^{\prime}}\le0\\
\Delta E_{kl,k^{\prime}l^{\prime}}>0 & ,
\end{array}\right.\label{eq:6}
\end{equation}
where $\Delta E_{kl,k^{\prime}l^{\prime}}$ is the change in the energy
after flipping the spins $\sigma_{kl}$ and $\sigma_{k^{\prime}l^{\prime}}$.
Thus, in this process, the new state is just accepted if $\Delta E_{kl,k^{\prime}l^{\prime}}>0$.

Repeating the Markov process $N$ times, we have one Monte Carlo Step
(MCS). In our simulations, for $p\ne0$, we have waited for $2\times10^{4}$
MCS for the system to reach the stationary state, for all the lattice
sizes. We used more $5\times10^{3}$ MCS to calculate the thermal
averages of the quantities of interest. The average over the samples
was done using $25$ independent samples for any lattice. On the other
hand, for the case $p=0$, we needed to wait for $5\times10^{5}$
MCS to reach the equilibrium state, and after $3\times10^{5}$ MCS
to calculate the thermal average, only over one sample.

The measured thermodynamic quantities in our simulations are: the
total magnetization per spin $\textrm{m}_{\textrm{L}}^{\textrm{F}}$,
the staggered magnetization per spin $\textrm{m}_{\textrm{L}}^{\textrm{AF}}$,
the magnetic susceptibility $\chi_{\textrm{L}}$ and the reduced fourth-order
Binder cumulant $\textrm{U}_{\textrm{L}}$:

\begin{equation}
\textrm{\ensuremath{\textrm{m}_{\textrm{L}}^{\textrm{F}}}}=\frac{1}{N}\left[\left\langle \sum_{kl}\sigma_{kl}\right\rangle \right],\label{eq:7}
\end{equation}

\begin{equation}
\textrm{m}_{\textrm{L}}^{\textrm{AF}}=\frac{1}{N}\left[\left\langle \sum_{kl}(-1)^{(k+l)}\sigma_{kl}\right\rangle \right],\label{eq:8}
\end{equation}

\begin{equation}
\chi_{\textrm{L}}=\frac{N}{k_{B}T}\left[\left\langle m^{2}\right\rangle -\left\langle m\right\rangle ^{2}\right],\label{eq:9}
\end{equation}

\begin{equation}
\textrm{U}_{\textrm{L}}=1-\frac{\left[\left\langle m^{4}\right\rangle \right]}{3\left[\left\langle m^{2}\right\rangle ^{2}\right]},\label{eq:10}
\end{equation}
where $\left[\ldots\right]$ denotes the average over the samples,
$\left\langle \ldots\right\rangle $ is the thermal average over the
MCS in the stationary state, and $m$ can be $\textrm{\ensuremath{\textrm{m}_{\textrm{L}}^{\textrm{F}}}}$
or $\textrm{\ensuremath{\textrm{m}_{\textrm{L}}^{\textrm{AF}}}}$
in Eq. (\ref{eq:9}) and (\ref{eq:10}), respectively. The lattice
sizes from $L=24$ to $L=256$ are simulated and the data are analyzed
via finite-size scaling theory (FSS). These Eqs. (\ref{eq:7}), (\ref{eq:8}),
(\ref{eq:9}) and (\ref{eq:10}) obey the following FSS relations
in the neighborhood of the stationary critical point $\lambda_{C}$:

\begin{equation}
m=L^{-\beta/\nu}m_{0}(L^{1/\nu}\varepsilon),\label{eq:11}
\end{equation}

\begin{equation}
\chi_{\textrm{L}}=L^{\gamma/\nu}\mathcal{X}_{0}(L^{1/\nu}\varepsilon),\label{eq:12}
\end{equation}

\begin{equation}
\textrm{U}_{\textrm{L}}=U_{0}(L^{1/\nu}\varepsilon),\label{eq:13}
\end{equation}
where $\varepsilon=(\lambda-\lambda_{C})/\lambda_{C}$, $\lambda$
can be $T$ or $q$. Here $m_{0}$, $\chi_{0}$ and $\textrm{U}_{0}$
are scaling functions, where $\beta$, $\gamma$, and $\nu$ are the
critical exponents related to magnetization, susceptibility, and the
length correlation, respectively. The derivative of Eq. (\ref{eq:13})
with respect to the parameter $\lambda$ gives us the following scaling
relation:

\begin{equation}
\textrm{U}'_{\textrm{L}}=\frac{L^{1/\nu}}{\lambda_{C}}U'_{0}(L^{1/\nu}\varepsilon).\label{eq:14}
\end{equation}

We have determined the critical exponent relations $\beta/\nu$, $\gamma/\nu$
and $\nu$ from slope of a log-log plot of $\textrm{m}_{\textrm{L}}(\lambda_{C})$,
$\mathcal{X}_{\textrm{L}}(\lambda_{C})$ or $\textrm{U}'_{\textrm{L}}(\lambda_{C})$
versus lattice size $L$ respectively. We also have used another alternative
method to estimate the values of the critical exponents, the data
collapse from the scaling functions.

\begin{figure}[th]
\begin{centering}
\includegraphics[clip,scale=0.33]{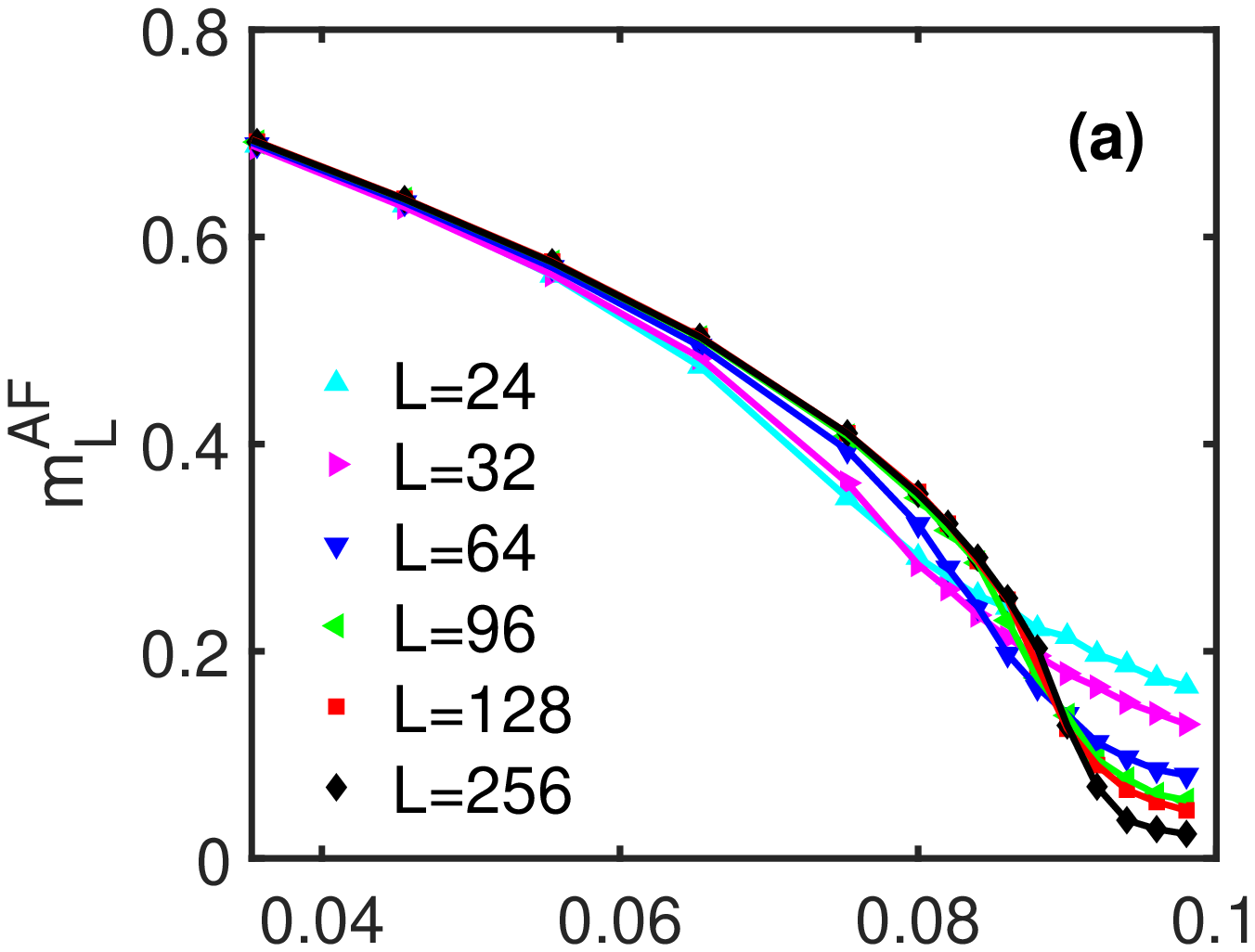}\includegraphics[clip,scale=0.33]{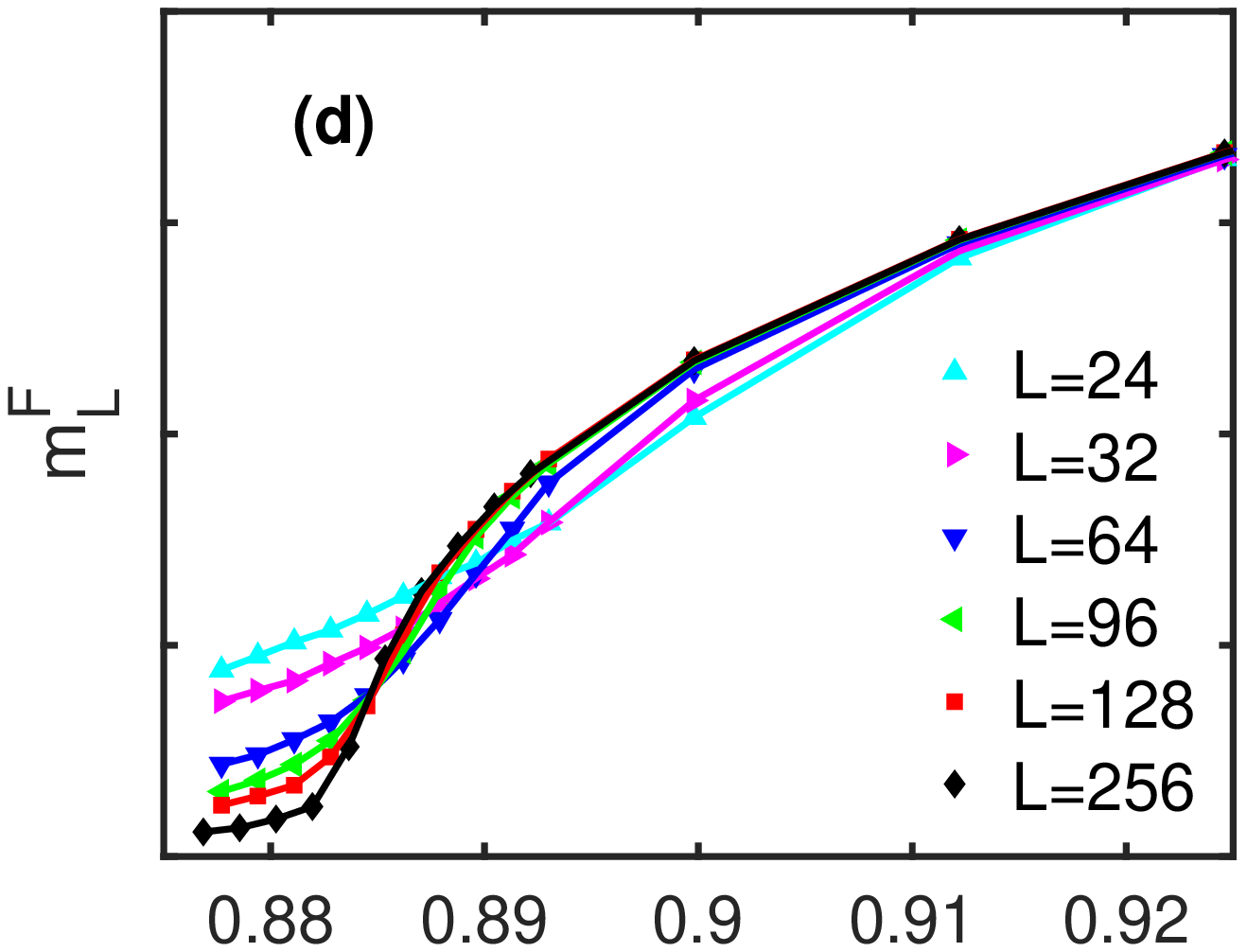}
\par\end{centering}
\begin{centering}
\includegraphics[clip,scale=0.33]{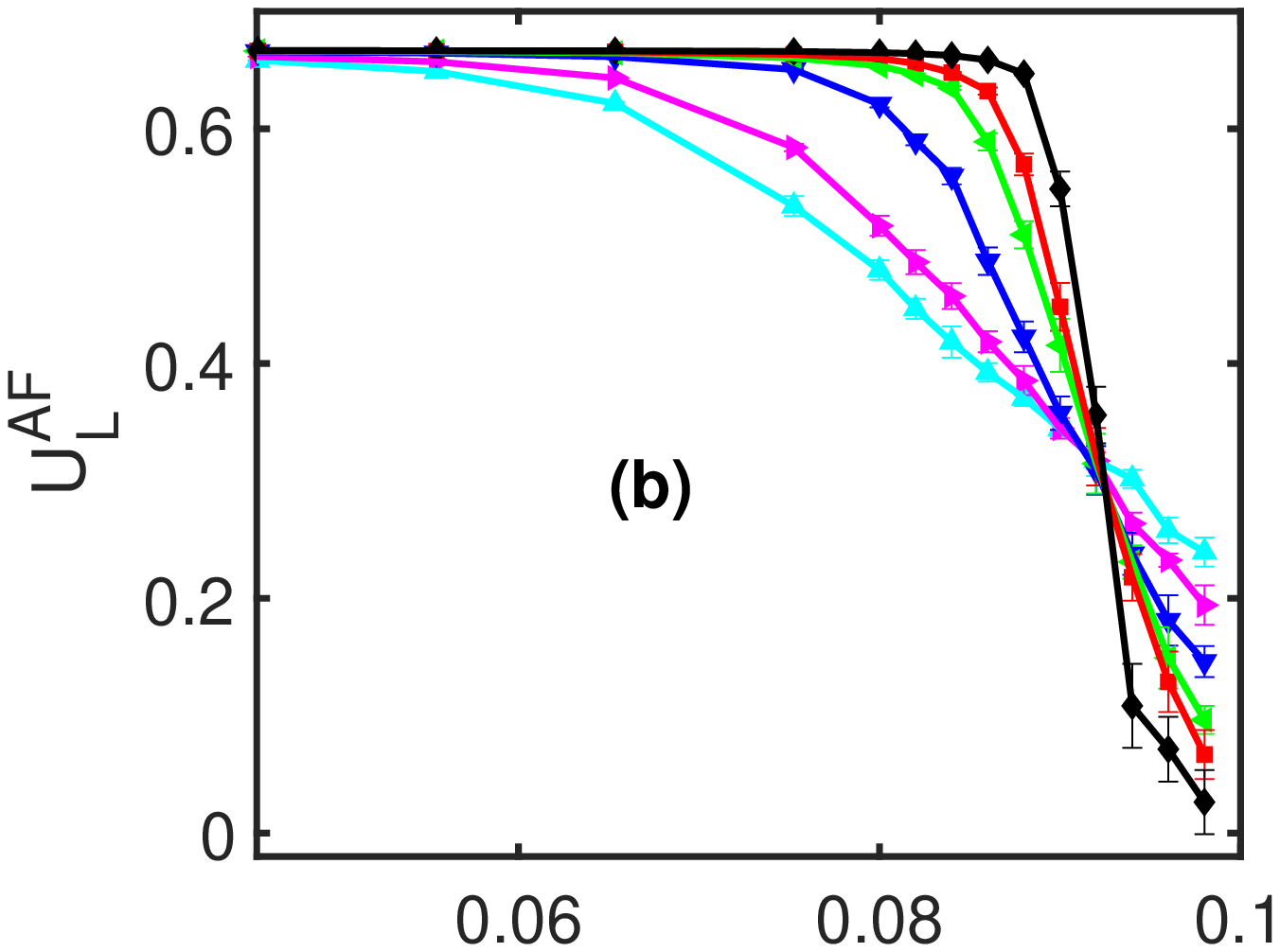}\includegraphics[clip,scale=0.33]{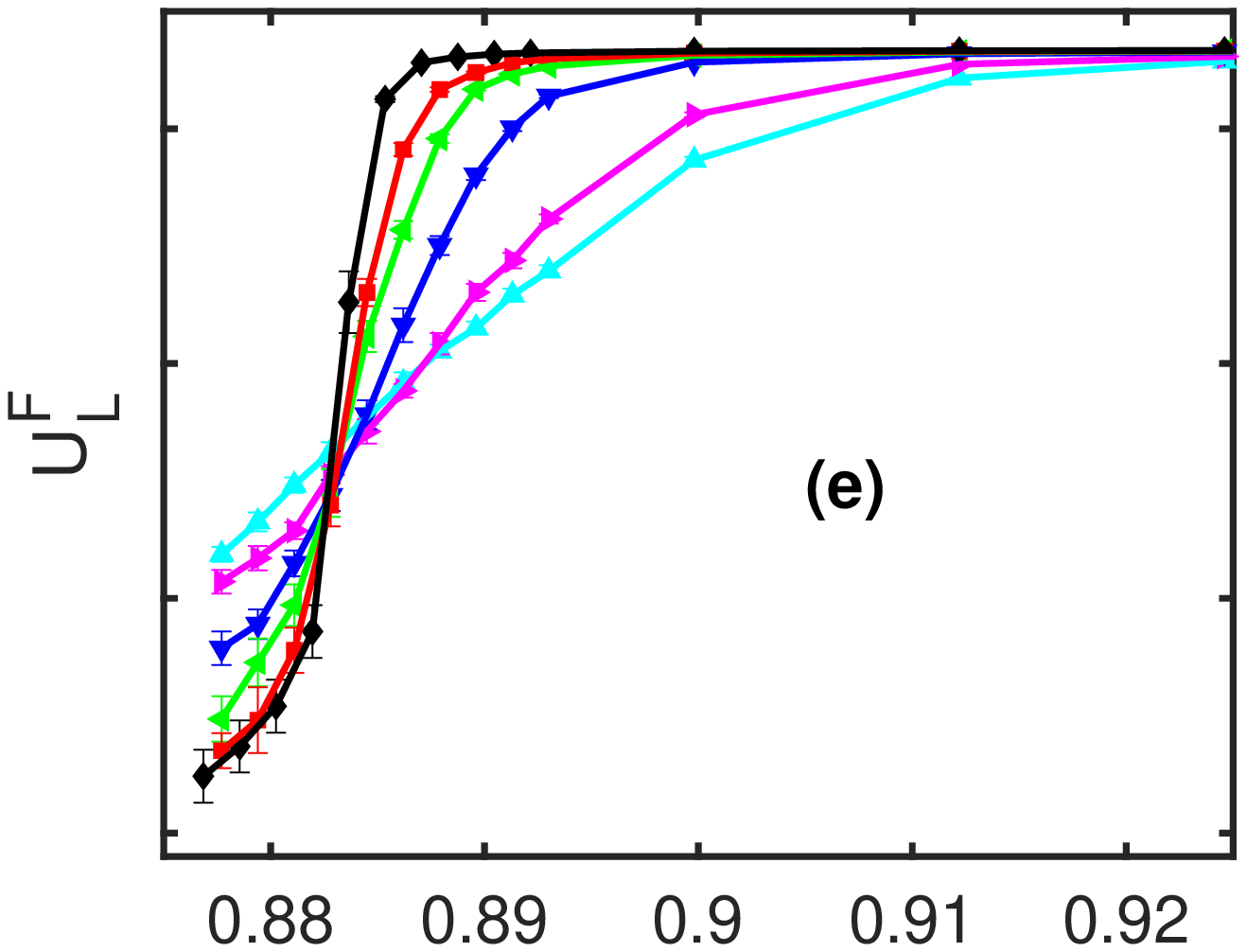}
\par\end{centering}
\begin{centering}
\includegraphics[clip,scale=0.33]{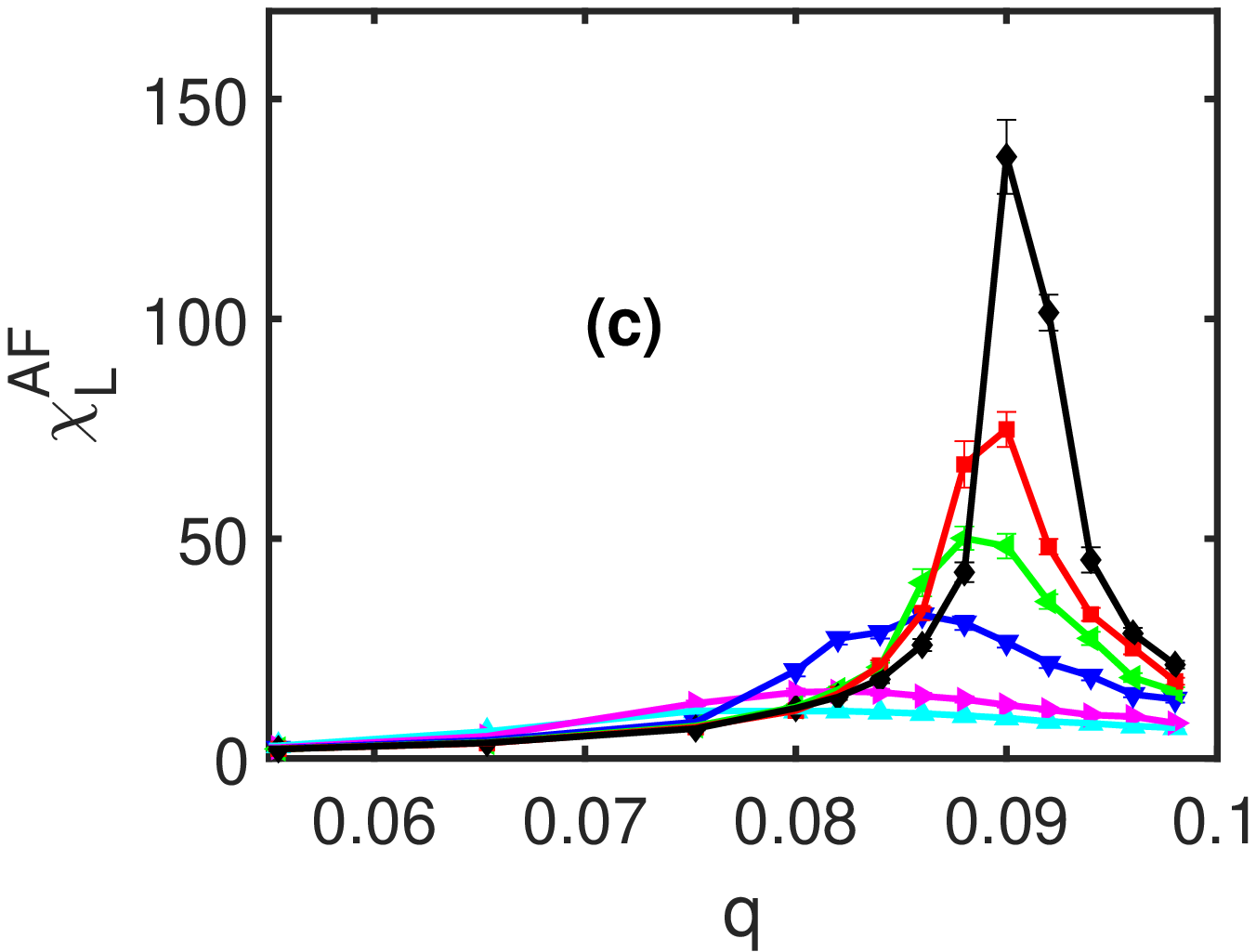}\includegraphics[clip,scale=0.33]{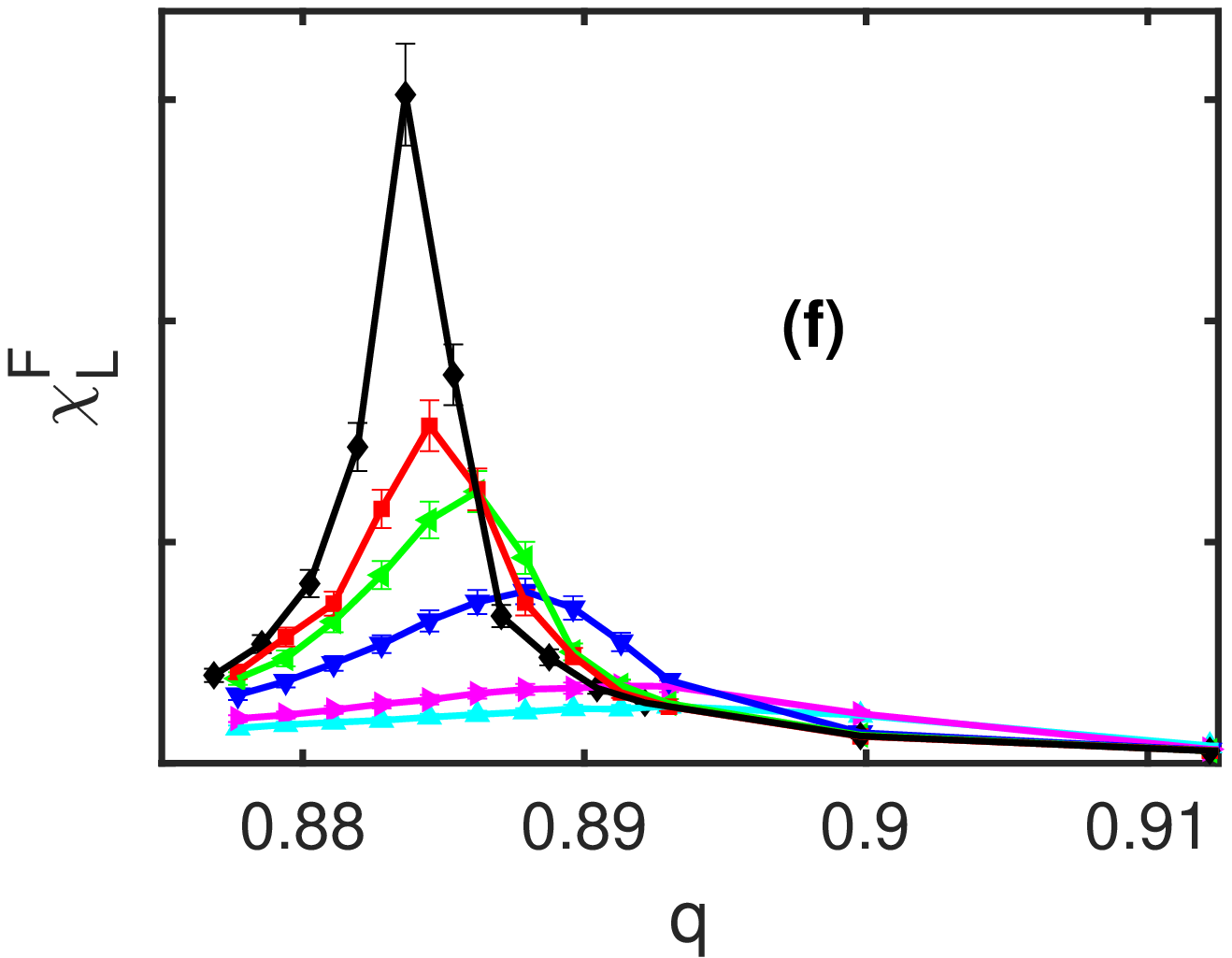}
\par\end{centering}
\caption{{\footnotesize{}Thermodynamic quantities in the phase transitions
of the non-equilibrium system, with $p=0.75$, $T=1$, and lattice
sizes $L$ shown in the figures. (a) Staggered magnetization $\textrm{m}_{\textrm{L}}^{\textrm{AF}}$,
in (b) and (c) we have respectively the Binder cumulant $\textrm{U}_{\textrm{L}}^{\textrm{AF}}$
and the staggered susceptibility $\chi_{\textrm{L}}^{\textrm{ AF}}$.
(d) Total magnetization $\textrm{m}_{\textrm{L}}^{\textrm{F}}$ of
the system is represented, the Binder cumulant $\textrm{U}_{\textrm{L}}^{\textrm{F}}$
in (e), and the total susceptibility $\chi_{\textrm{L }}^{\textrm{F}}$
in (f). The error bars in the magnetization are smaller than the size
of the symbols, so for a better interpretation of the results, these
were omitted. \label{fig:2}}}
\end{figure}

\subsection{Results and Discussions\label{subsec:Results}}

In this section, we illustrate and discuss the results of the magnetic
properties of the Ising model on a 2D A-SWN at the non-equilibrium
regime by the two competing dynamics. For the study about the critical
behavior and phase transitions at the non-equilibrium system, it was
convenient to fix the temperature $T$, additive probability $p$,
and to use the competition parameter $q$ as a variable to transit
between the ordered to disordered phases in the regions of $T$ and
$p$ of the phase diagram. It is convenient because the two-spin flip
mechanism is independent of the temperature $T$, and in the present
work we do not have used $p$ as a variable to identify the phase
transitions.

Before studying the thermal phase diagrams, we will present the best
results for the behavior of thermodynamic quantities and critical
point values. These results were obtained where most sites have the
same coordination number $z=5$. Therefore, in Fig. \ref{fig:2},
we have shown one of the best results for the thermodynamic quantities
obtained in the stationary state as a function of $q$, for fixed
$p=0.75$ and $T=1$. We can see the self-organization in the system,
by finding an $AF$ phase, being represented in the staggered magnetization
$\textrm{m}_{\textrm{L}}^{\textrm{AF}}$. These because in high values
of $q$ we have the transition between the $F$ to $P$ phase (see
Fig. \ref{fig:2}(d)) and from this $P$ phase to the ordered $AF$
phase (see Fig. \ref{fig:2}(a)) as we increase the flow of energy
into the system ($q\rightarrow0)$. For these magnetizations, we also
have their respective reduced fourth-order Binder cumulants, $\textrm{U}_{\textrm{L}}^{\textrm{AF}}$
(Fig. \ref{fig:2}(b)) and $\textrm{U}_{\textrm{L}}^{\textrm{F}}$
(Fig. \ref{fig:2}(e)) beyond the magnetic susceptibilities $\chi_{\textrm{L}}^{\textrm{AF}}$
(Figs. \ref{fig:2}(c)) and $\chi_{\textrm{L}}^{\textrm{F}}$ (Fig.
\ref{fig:2}(f)).
\begin{center}
\begin{figure}
\begin{centering}
\includegraphics[clip,scale=0.33]{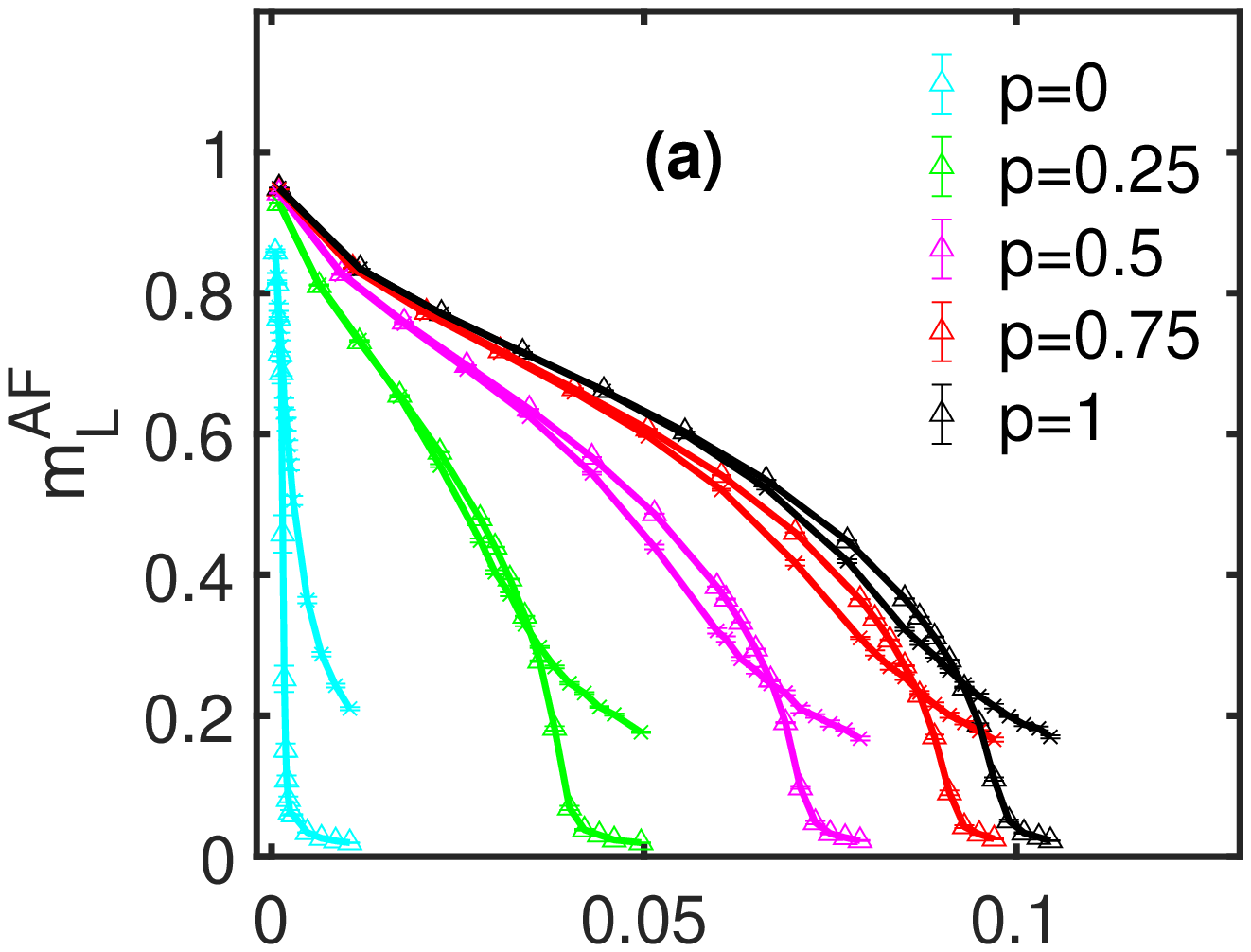}\includegraphics[clip,scale=0.33]{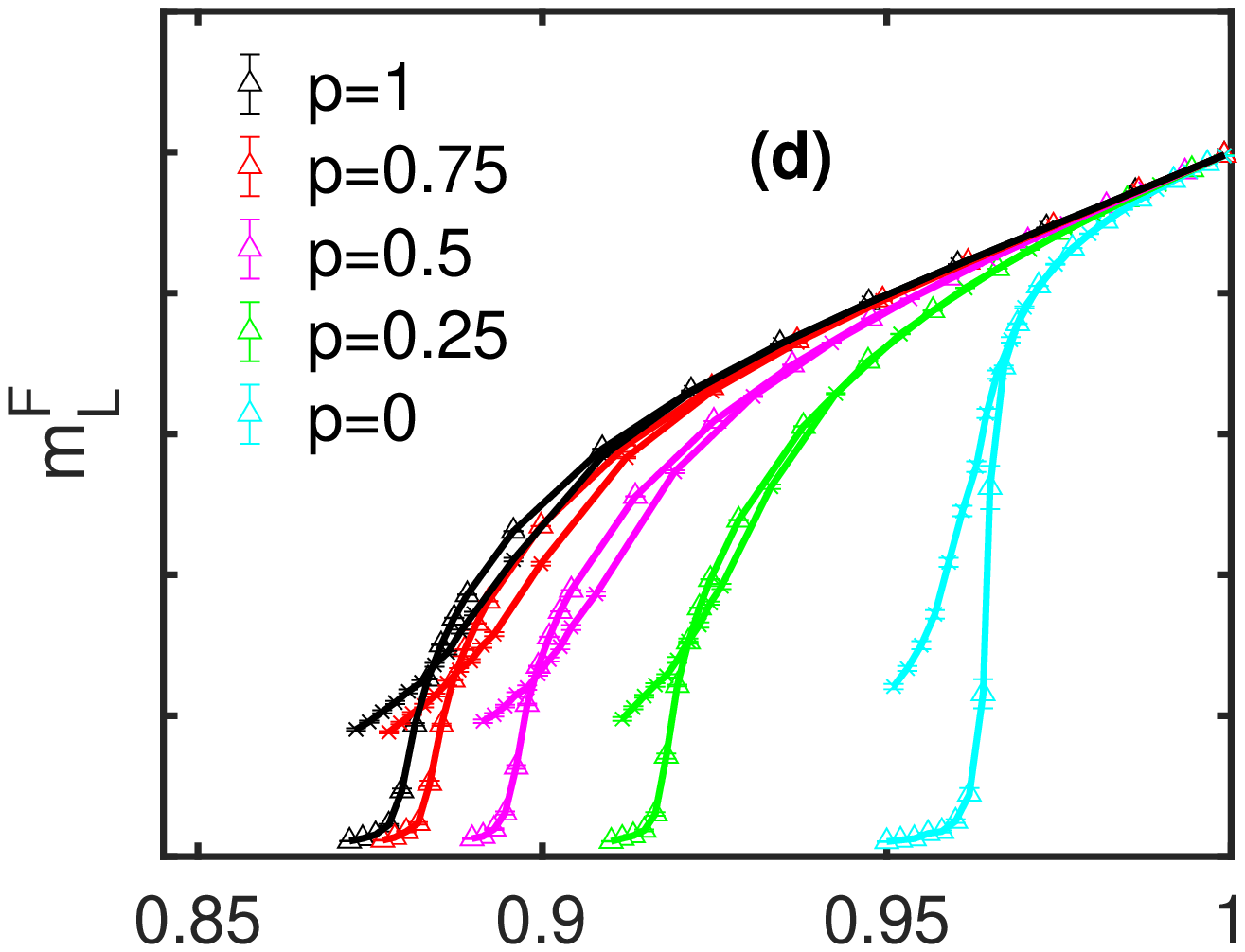}
\par\end{centering}
\begin{centering}
\includegraphics[clip,scale=0.33]{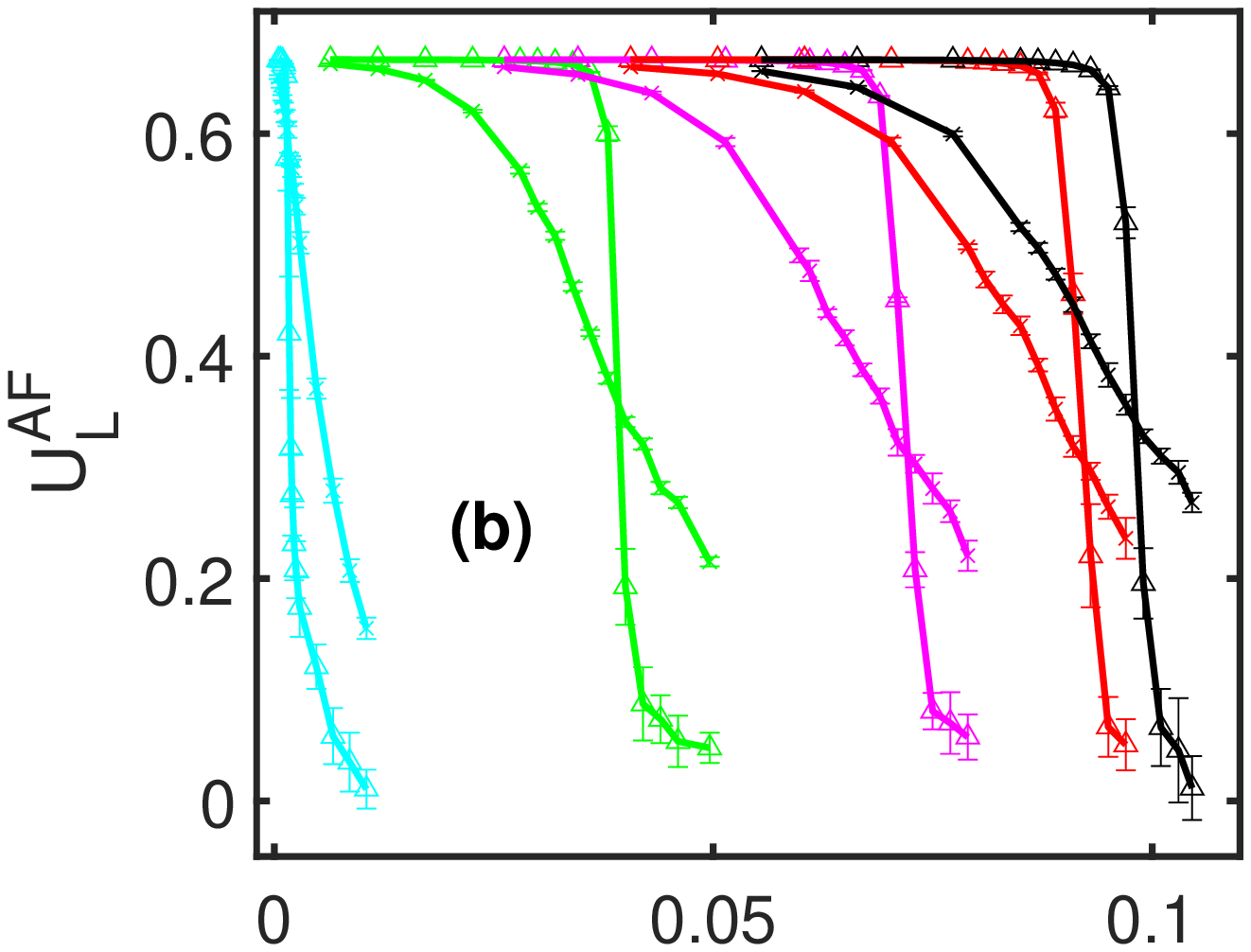}\includegraphics[clip,scale=0.33]{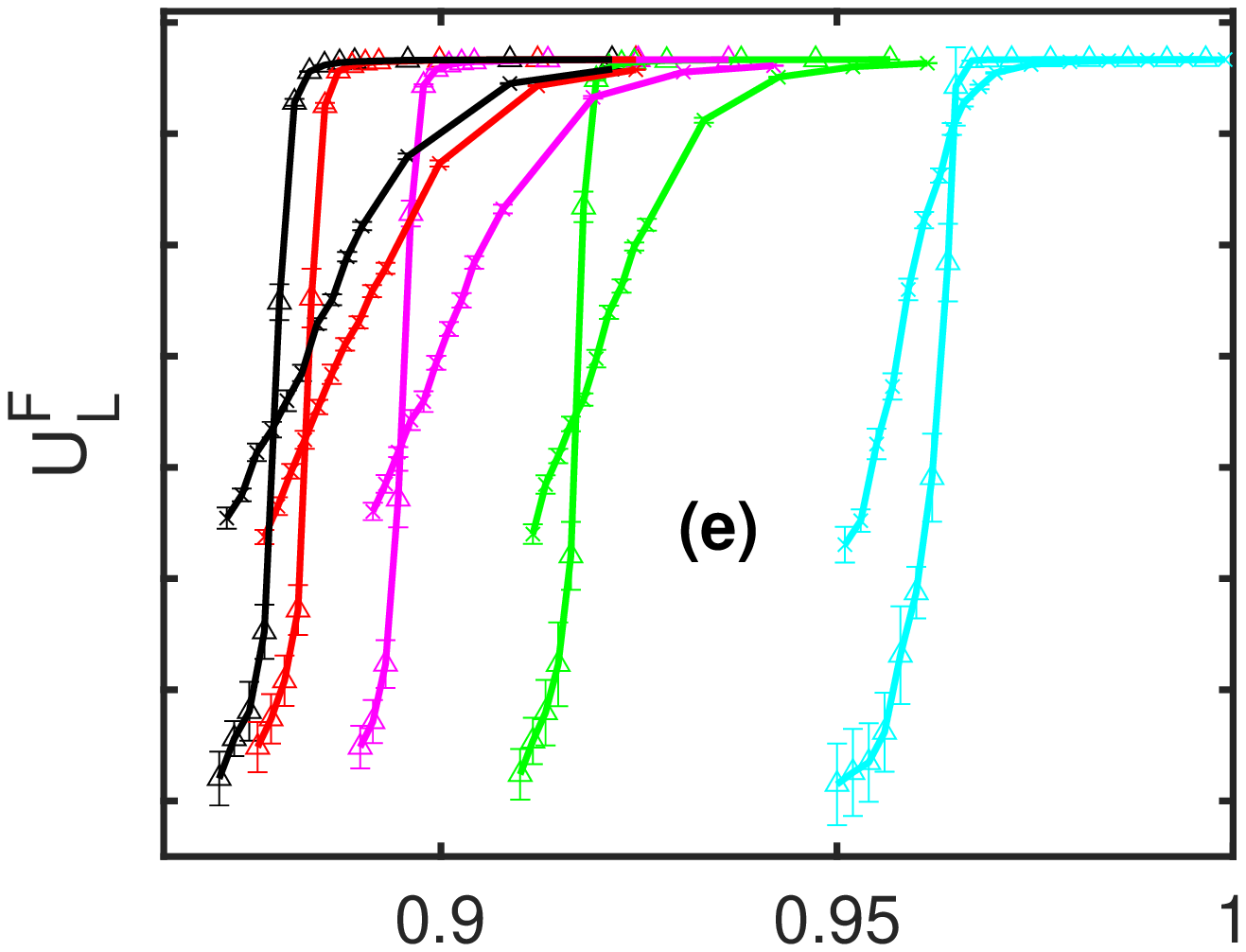}
\par\end{centering}
\begin{centering}
\includegraphics[clip,scale=0.33]{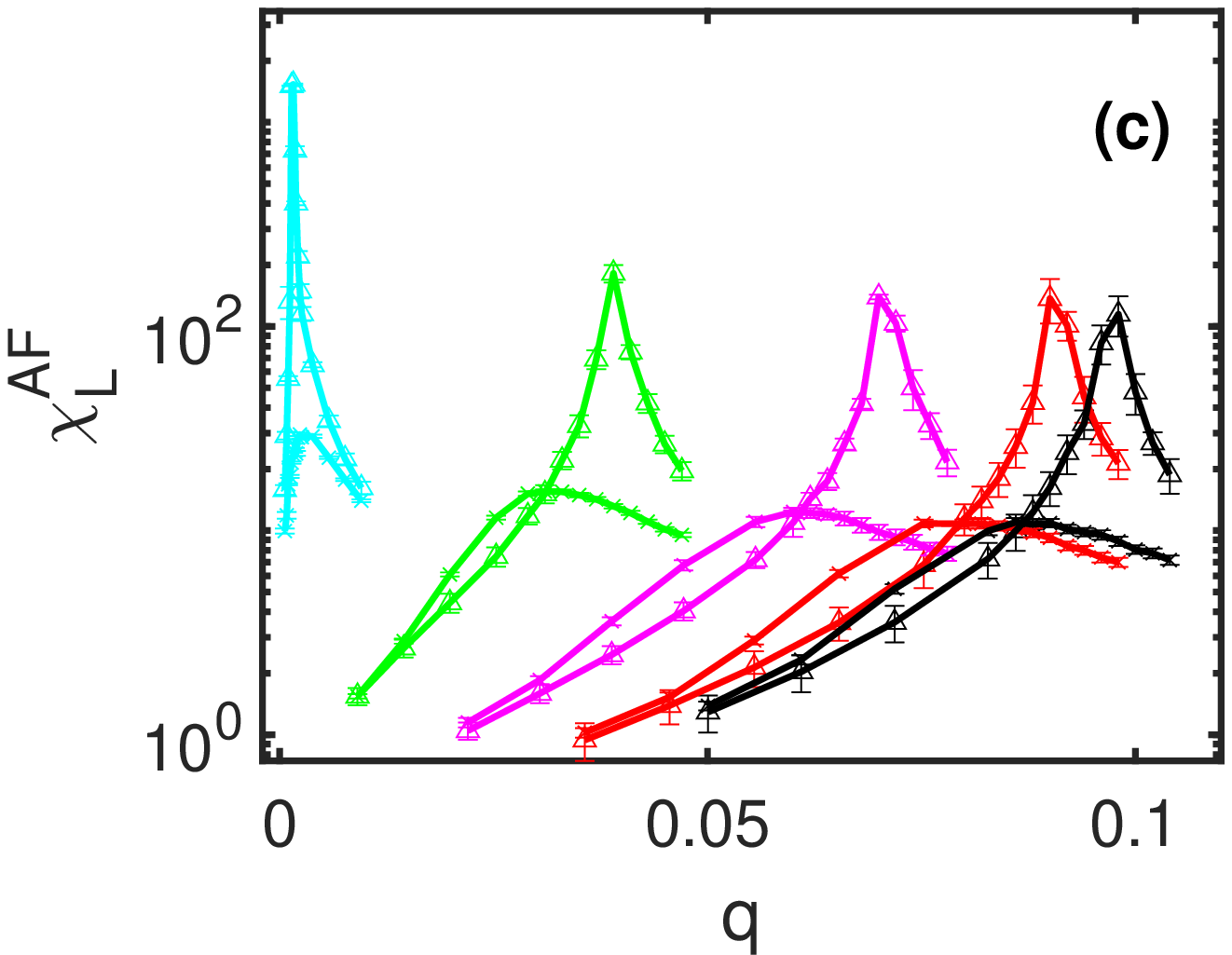}\includegraphics[clip,scale=0.33]{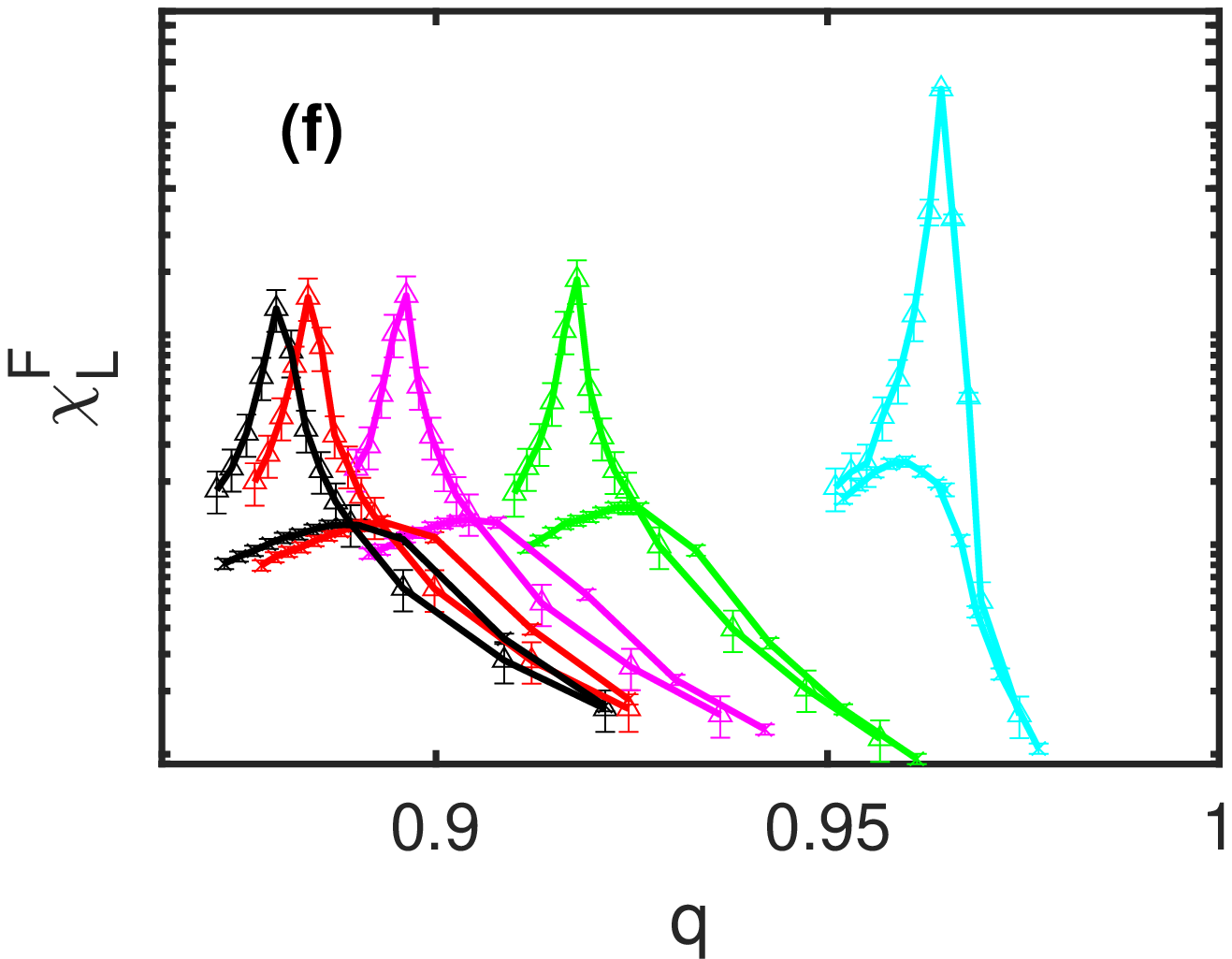}
\par\end{centering}
\caption{{\footnotesize{}Thermodynamic quantities as a function of $q$, for
different values of $p$ as indicated in the figures. We fixed the
value of $T=1$, and for lattice sizes $L=256\,\,(\triangle)$ and
$L=24\,\,(\times)$. (a) Staggered magnetization $\textrm{m}_{\textrm{L}}^{\textrm{AF}}$,
in (b) and (c) we have respectively the staggered susceptibility $\chi_{\textrm{L}}^{\textrm{ AF}}$
and Binder cumulative $\textrm{U}_{\textrm{L}}^{\textrm{AF}}$. (d)
Total magnetization $\textrm{m}_{\textrm{L}}^{\textrm{F}}$ of the
system is represented, the Binder cumulant $\textrm{U}_{\textrm{L}}^{\textrm{F}}$
in (e), and the total susceptibility $\chi_{\textrm{L }}^{\textrm{F}}$
in (f). \label{fig:3}}}
\end{figure}
\par\end{center}

The thermodynamic quantities for the other $p$ values in the A-SWN
regime, such as $p=0.25$, $p=0.5$, $p=0.75$, and $p=1$, have also
been computed. In the order to compare the behavior during phase transitions,
we also exhibited the same thermodynamic quantities in the conventional
square lattice Ising model, $p=0$, in Fig. \ref{fig:3}. These result
can see in details for the $\textrm{m}_{\textrm{L}}^{\textrm{AF}}$
and $\textrm{m}_{\textrm{L}}^{\textrm{F}}$ in Figs. \ref{fig:3}(a)
and \ref{fig:3}(d), respectively, $\textrm{U}_{\textrm{L}}^{\textrm{AF}}$
in Fig. \ref{fig:3}(b) and $\textrm{U}_{\textrm{L}}^{\textrm{F}}$
in Fig. \ref{fig:3}(e), in addition to $\chi_{\textrm{L}}^{\textrm{AF}}$
in Fig. \ref{fig:3}(c) and $\chi_{\textrm{L}}^{\textrm{F}}$ in Fig.
\ref{fig:3}(f). We have presented only the smaller $(L=24)$ and
the larger $(L=256)$ linear lattice size and they are enough so that
we can observe the finite-size behavior and the critical point change
$q_{c}$ as we increase $p$. On the other hand, for the calculation
of $q_{c}$, we have used all six lattice sizes of the system.
\begin{center}
\begin{figure}
\begin{centering}
\includegraphics[scale=0.5]{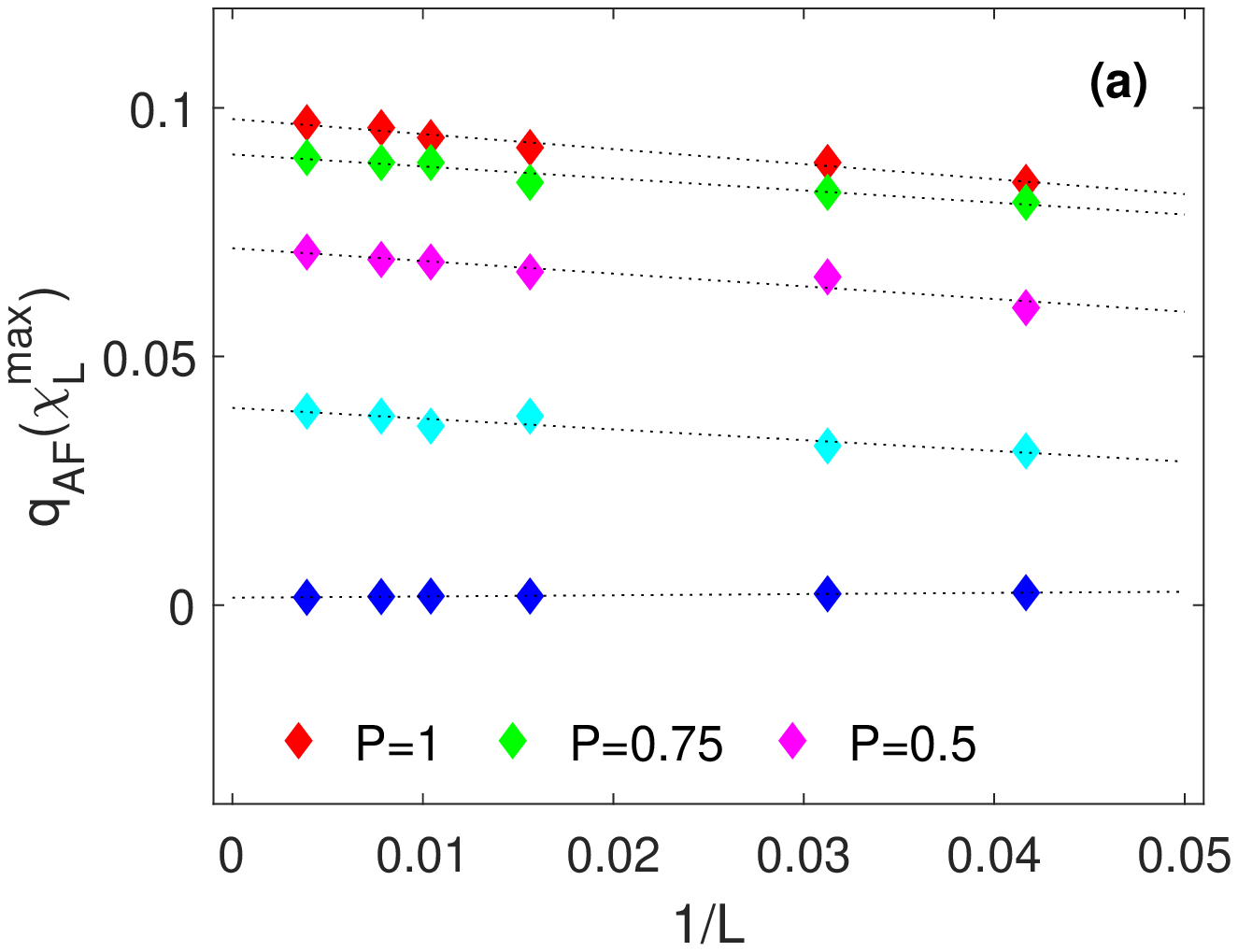}
\par\end{centering}
\begin{centering}
\vspace{0.1cm}
\par\end{centering}
\begin{centering}
\includegraphics[scale=0.5]{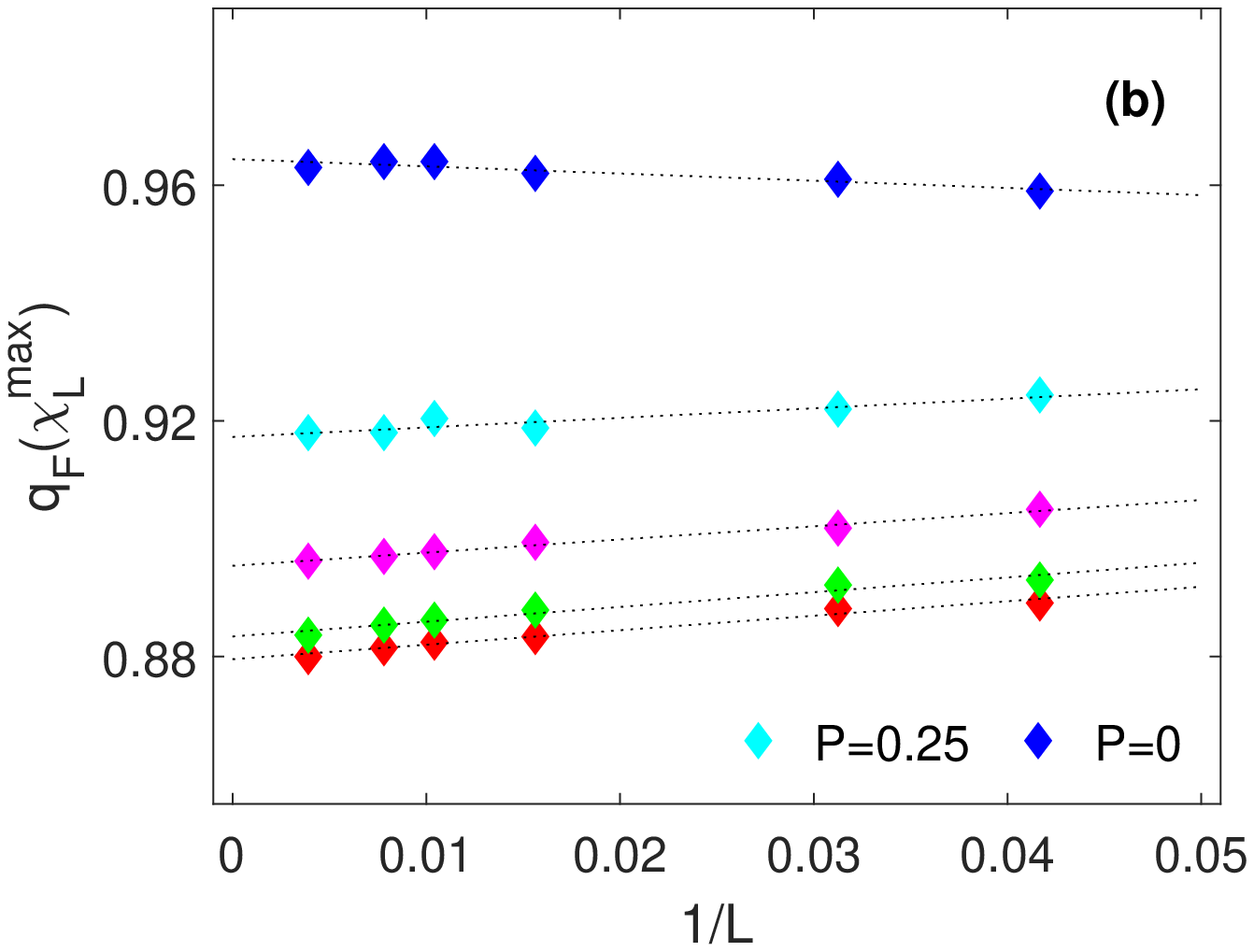}
\par\end{centering}
\caption{{\footnotesize{}Extrapolation of the critical transition probability
$q_{c}$ obtained for linear lattice sizes $24\le L\le256$, and for
different values of $p$ as indicated in the figures. (a) The points
in the transition between $AF-P$ phases are represented, and (b)
the points in the transition between the $P-F$ phases. The values
of critical transition probabilities $q_{c}(L\rightarrow\infty)$
can be seen in Table \ref{tab:1} and Table \ref{tab:2}, respectively.
\label{fig:4}}}
\end{figure}
\par\end{center}

To evaluate the $q_{c}$, we have employed two methods. Firstly, we
obtained by extrapolating the susceptibility discontinuity to when
$L\to\infty$, which returns $q_{c}(\infty)$, using finite lattice
sizes $24\le L\le256$, in the plot of maximum susceptibility as a
function of $1/L$. Secondly, we obtained by the crossing of the Binder
cumulant curves for the different lattice sizes $L$. In Fig. \ref{fig:4}
the values of $q$ are displayed where the susceptibility has its
maximum value, $\chi_{\textrm{L}}^{max}$, as a function of $1/L$
for the values of $p$ selected. We also have the best fit of the
points, which is a linear fit, and for the extrapolation, when $L\to\infty$,
we have the estimated of $q_{c}$ by using the linear coefficient,
i.e., we have made the infinite-size extrapolation in according to
$q(\chi_{\textrm{L}}^{max})-q_{c}(\infty)=\alpha L^{-1}$. By extrapolation,
the critical points $q_{c}(\infty)$ in the transition between the
$AF-P$ phases are represented in Fig. \ref{fig:4}(a) and the transition
between the $F-P$ phases are represented in Fig. \ref{fig:4}(b).

The critical point values using magnetic susceptibility data, $q_{c}^{\chi}$,
and their respective errors are exhibited in Table \ref{tab:1} for
the transition between the $AF-P$ phases. In this transition, we
also have used the crossing of the Binder cumulant curves in the selected
lattice sizes $24\le L\le256$, to obtain the another estimate for
the critical points, $q_{c}^{U}$, which are shown in Table \ref{fig:3},
and the characterization of the second-order phase transition in the
system \citep{28,29,30,31}. For the transition between $F-P$ phases,
the values of $q_{c}^{\chi}$ obtained are exhibited in Table \ref{tab:2},
and the critical points obtained through the crossing of the Binder
cumulant curves $q_{c}^{U}$, can be seen in Table \ref{tab:4}. The
critical points obtained in both methods are equivalent. 
\begin{center}
\begin{figure}
\begin{centering}
\includegraphics[bb=15bp 0bp 800bp 654bp,clip,scale=0.15]{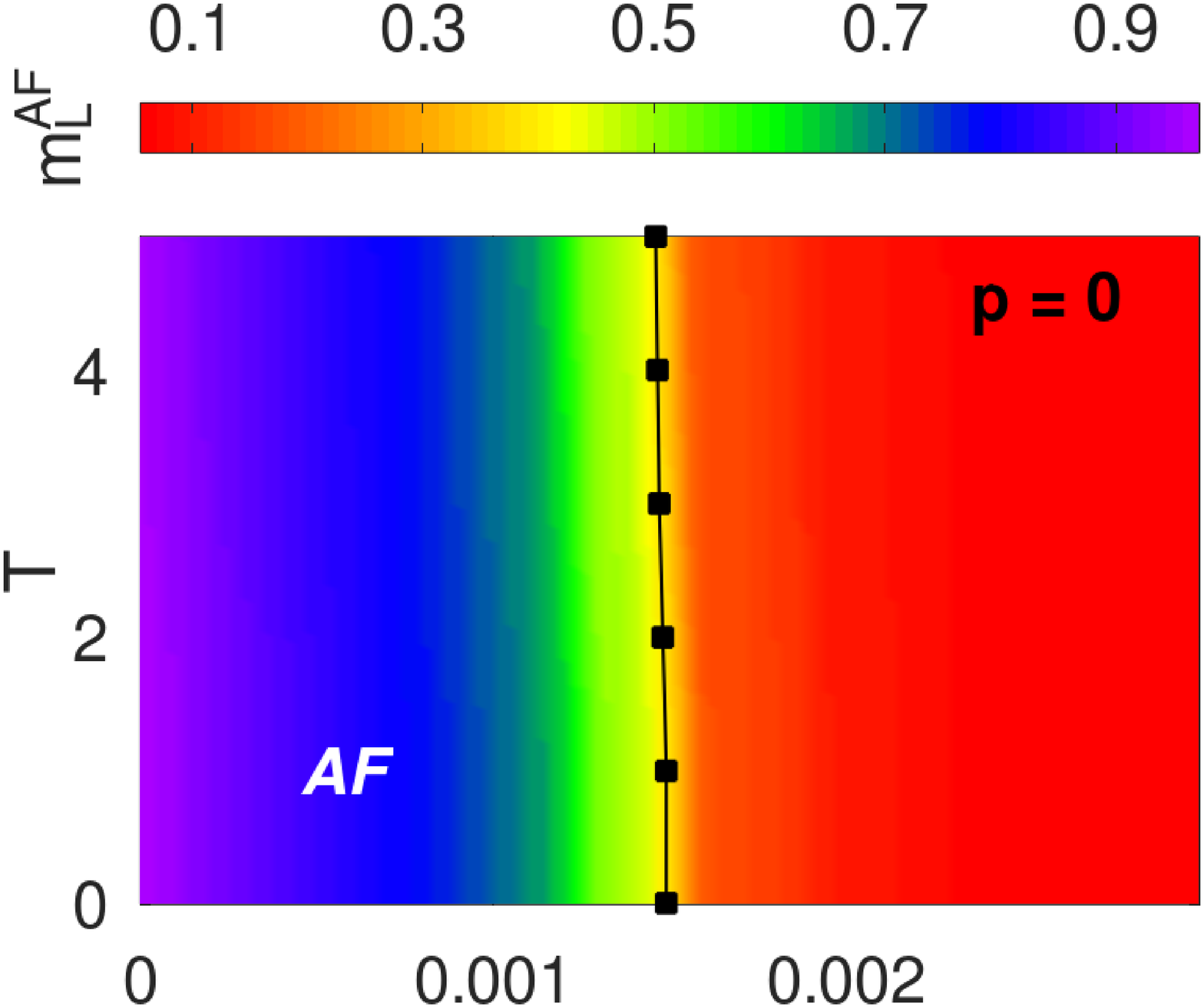}\includegraphics[bb=100bp 0bp 872bp 654bp,clip,scale=0.15]{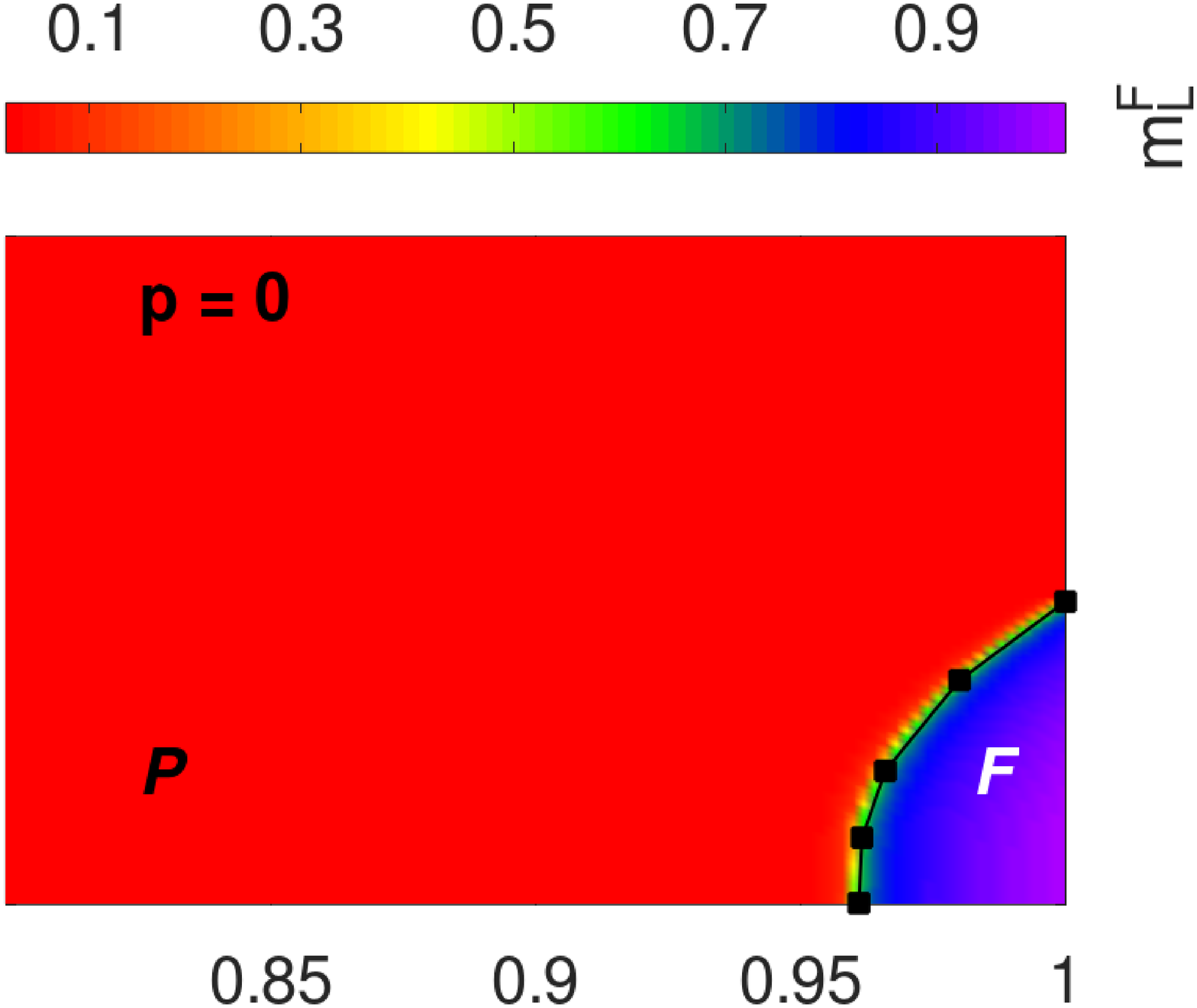}
\par\end{centering}
\begin{centering}
\includegraphics[bb=15bp 45bp 800bp 650bp,clip,scale=0.15]{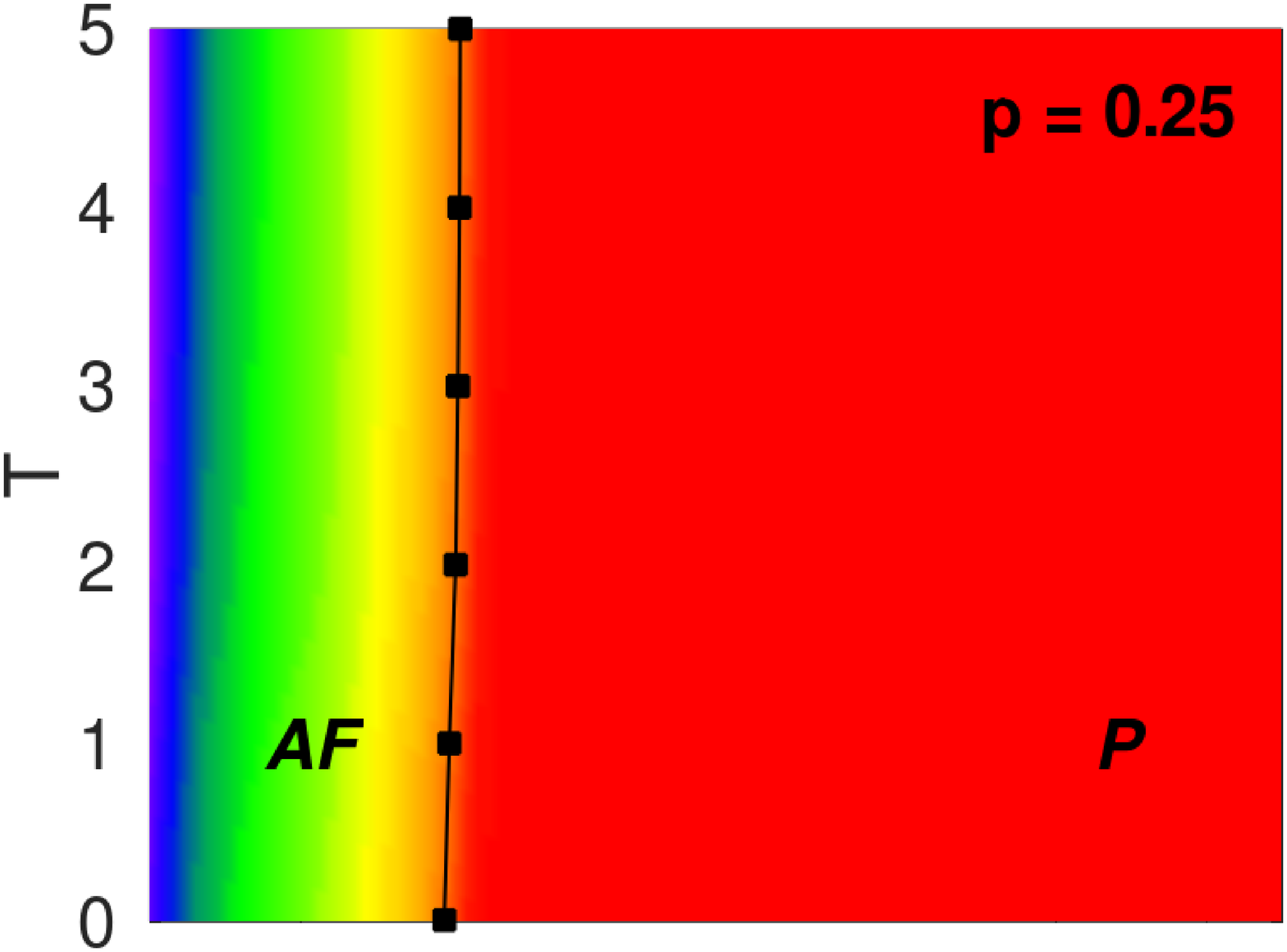}\includegraphics[bb=100bp 45bp 872bp 650bp,clip,scale=0.15]{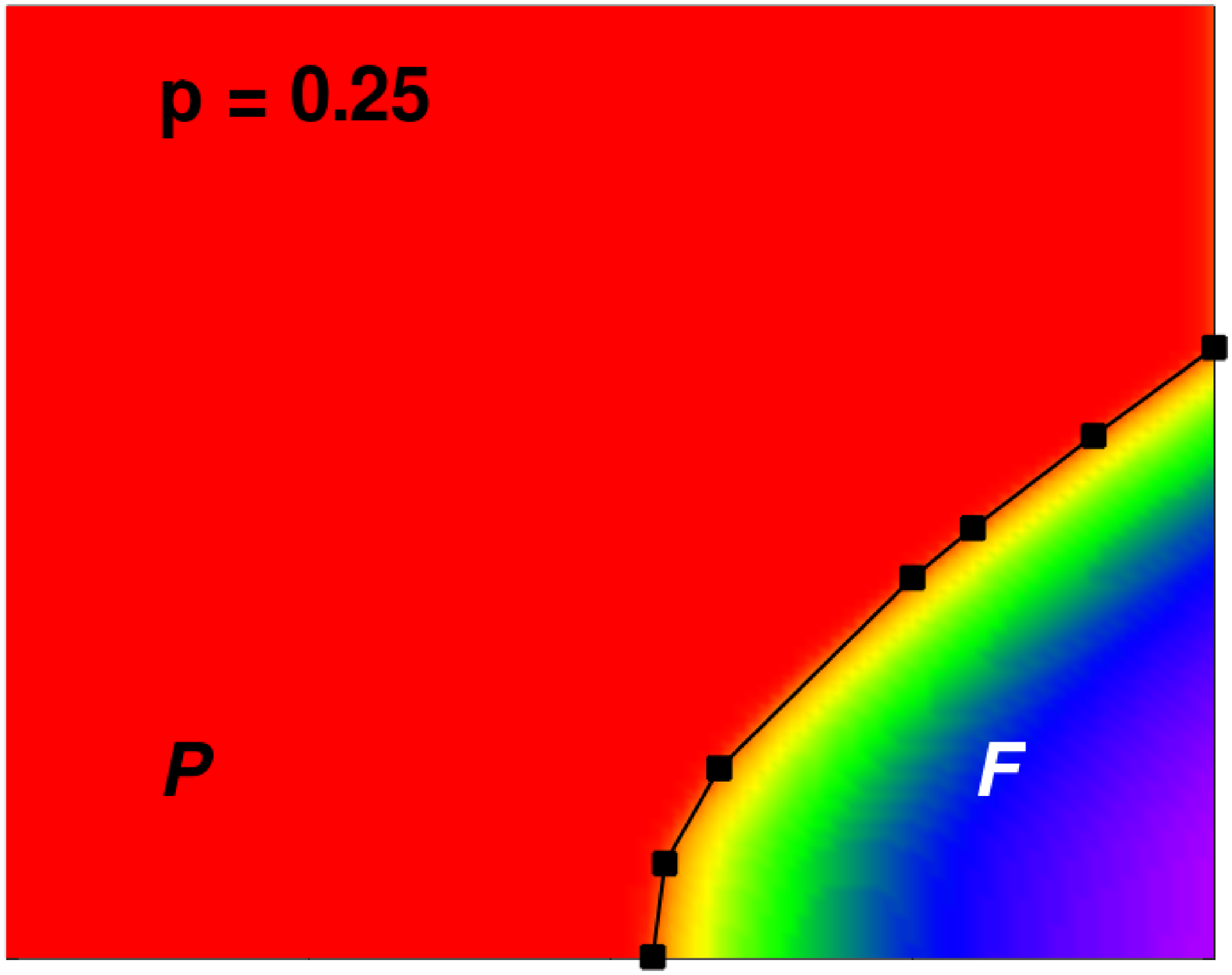}
\par\end{centering}
\begin{centering}
\includegraphics[bb=15bp 45bp 800bp 650bp,clip,scale=0.15]{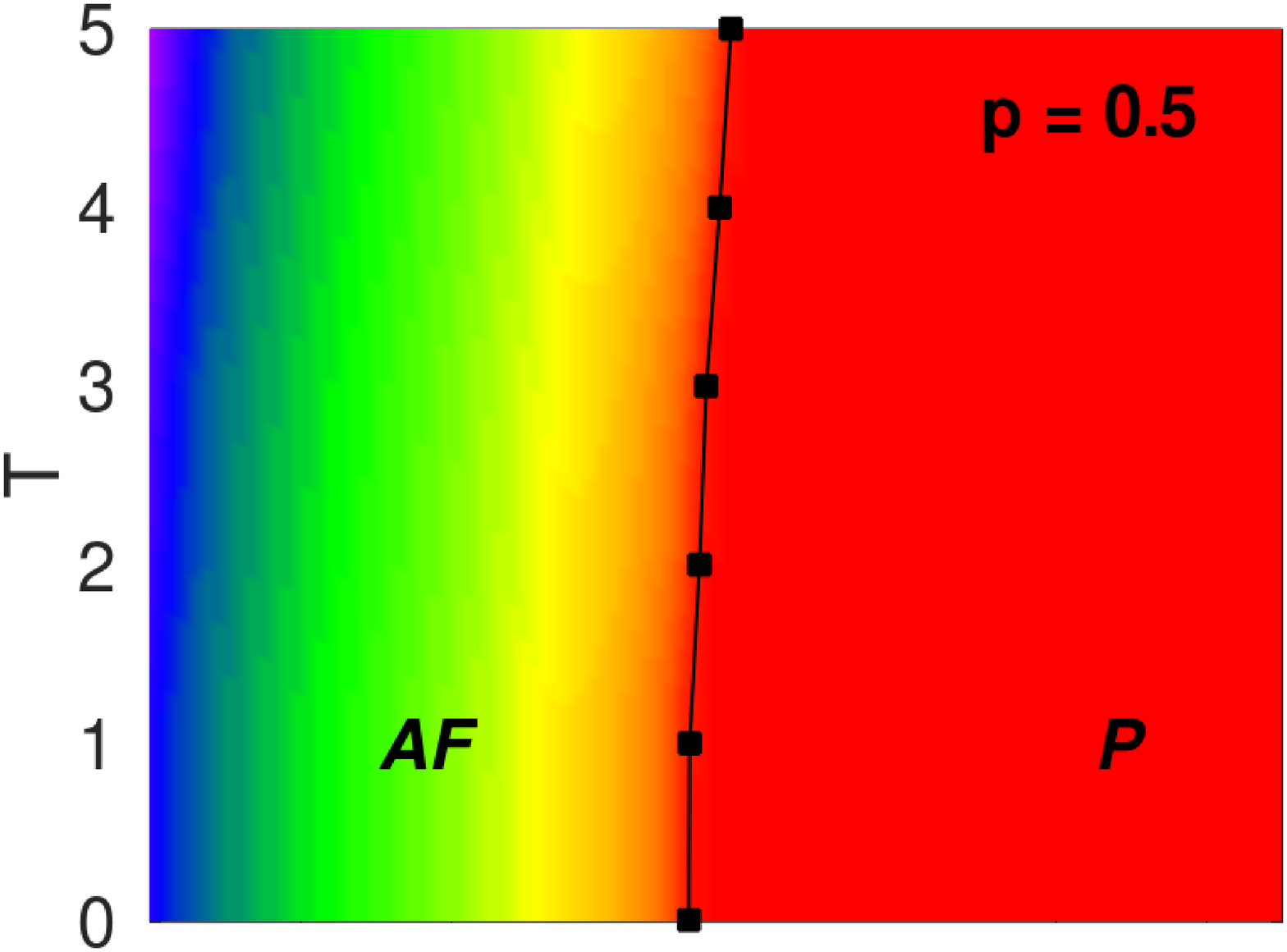}\includegraphics[bb=100bp 45bp 872bp 650bp,clip,scale=0.15]{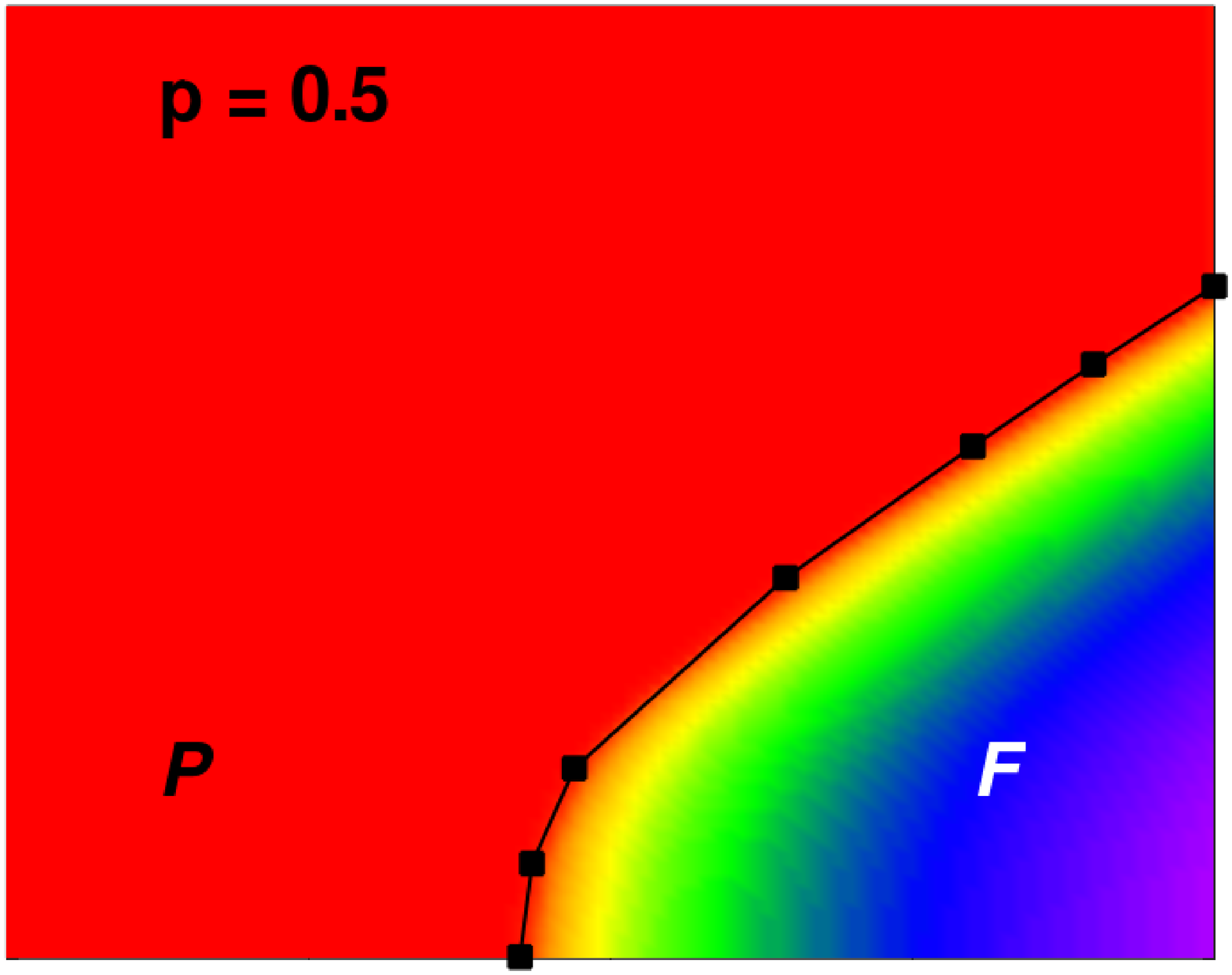}
\par\end{centering}
\begin{centering}
\includegraphics[bb=15bp 45bp 800bp 650bp,clip,scale=0.15]{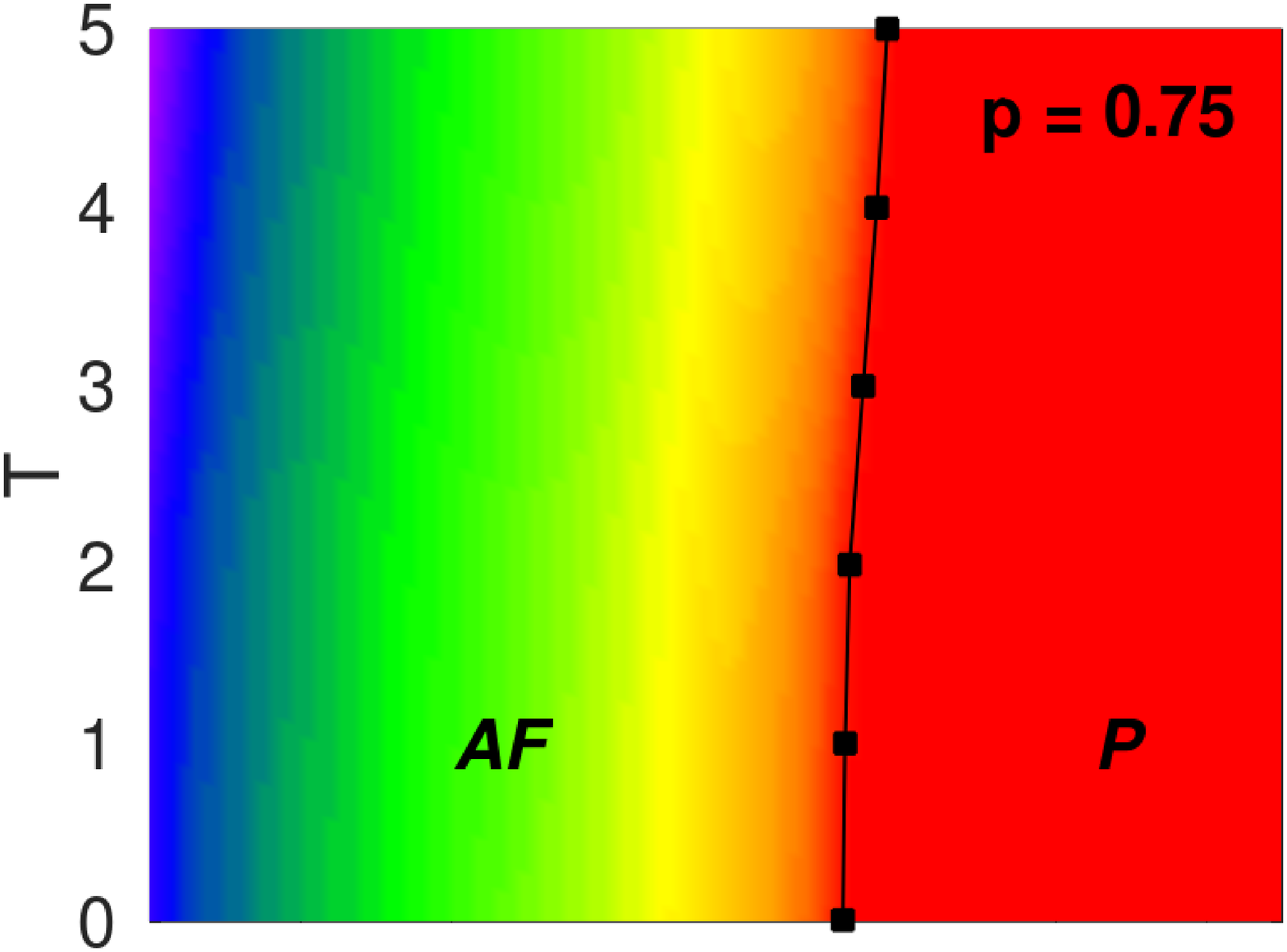}\includegraphics[bb=100bp 45bp 872bp 650bp,clip,scale=0.15]{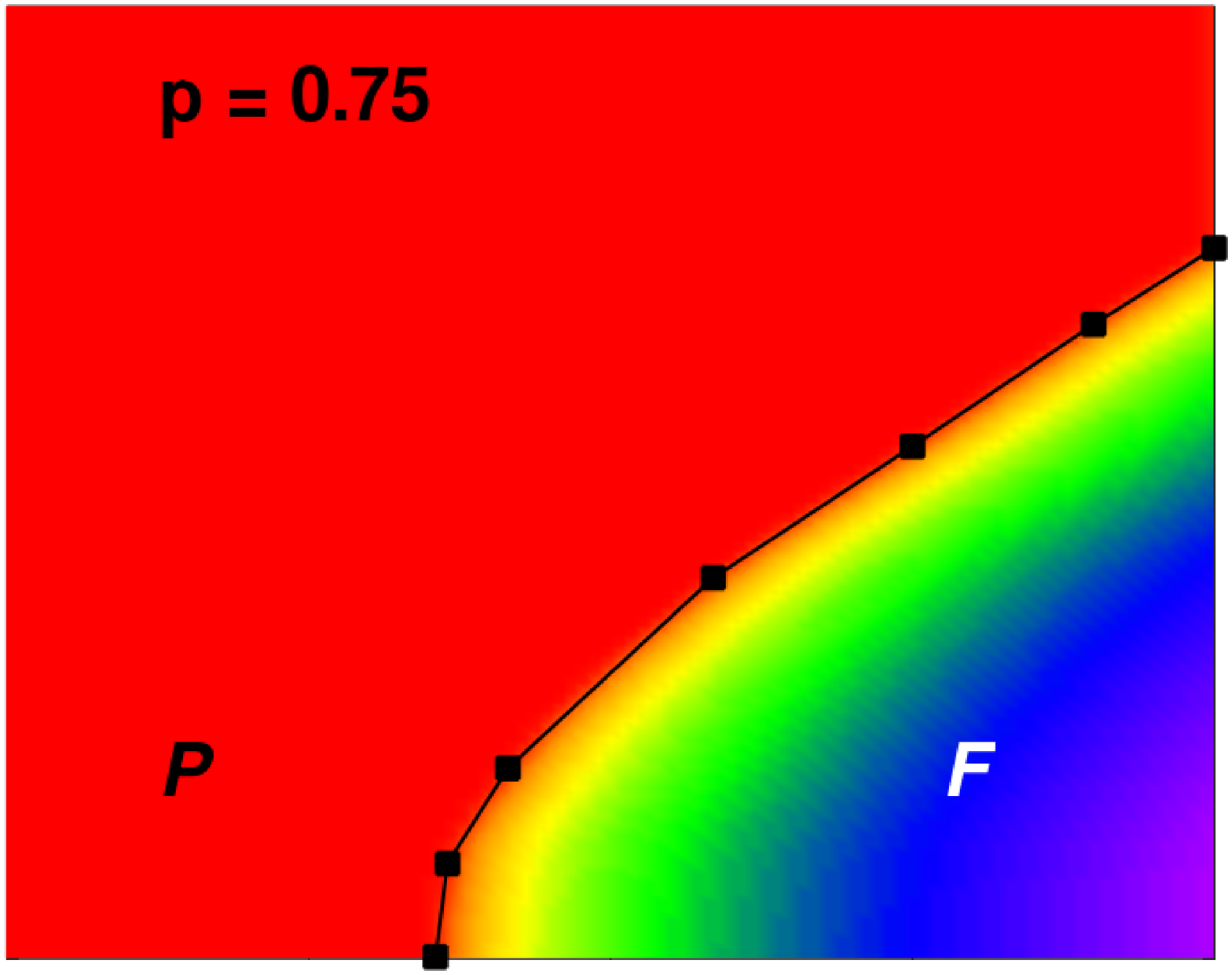}
\par\end{centering}
\begin{centering}
\includegraphics[bb=15bp 0bp 800bp 650bp,clip,scale=0.15]{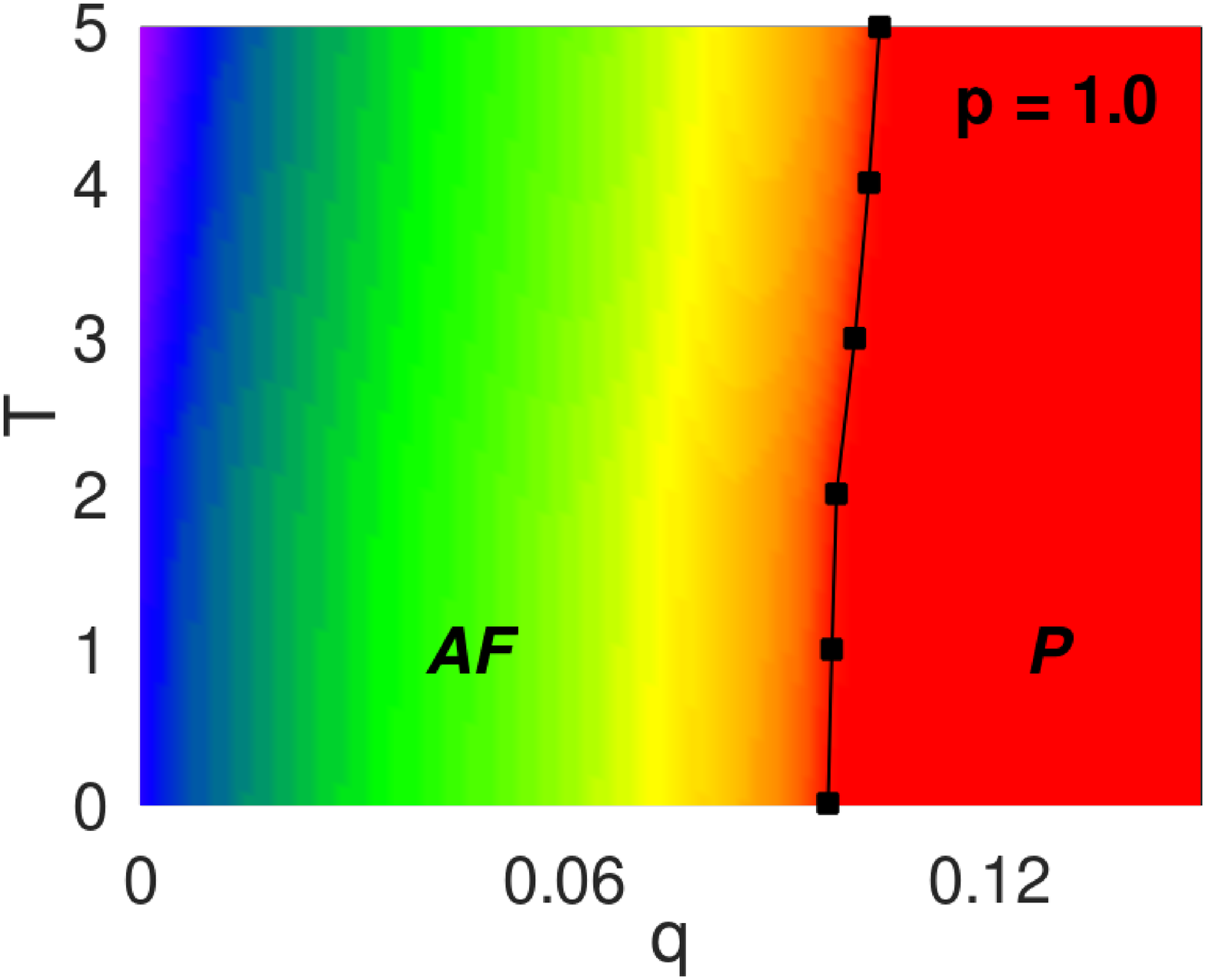}\includegraphics[bb=100bp 0bp 872bp 650bp,clip,scale=0.15]{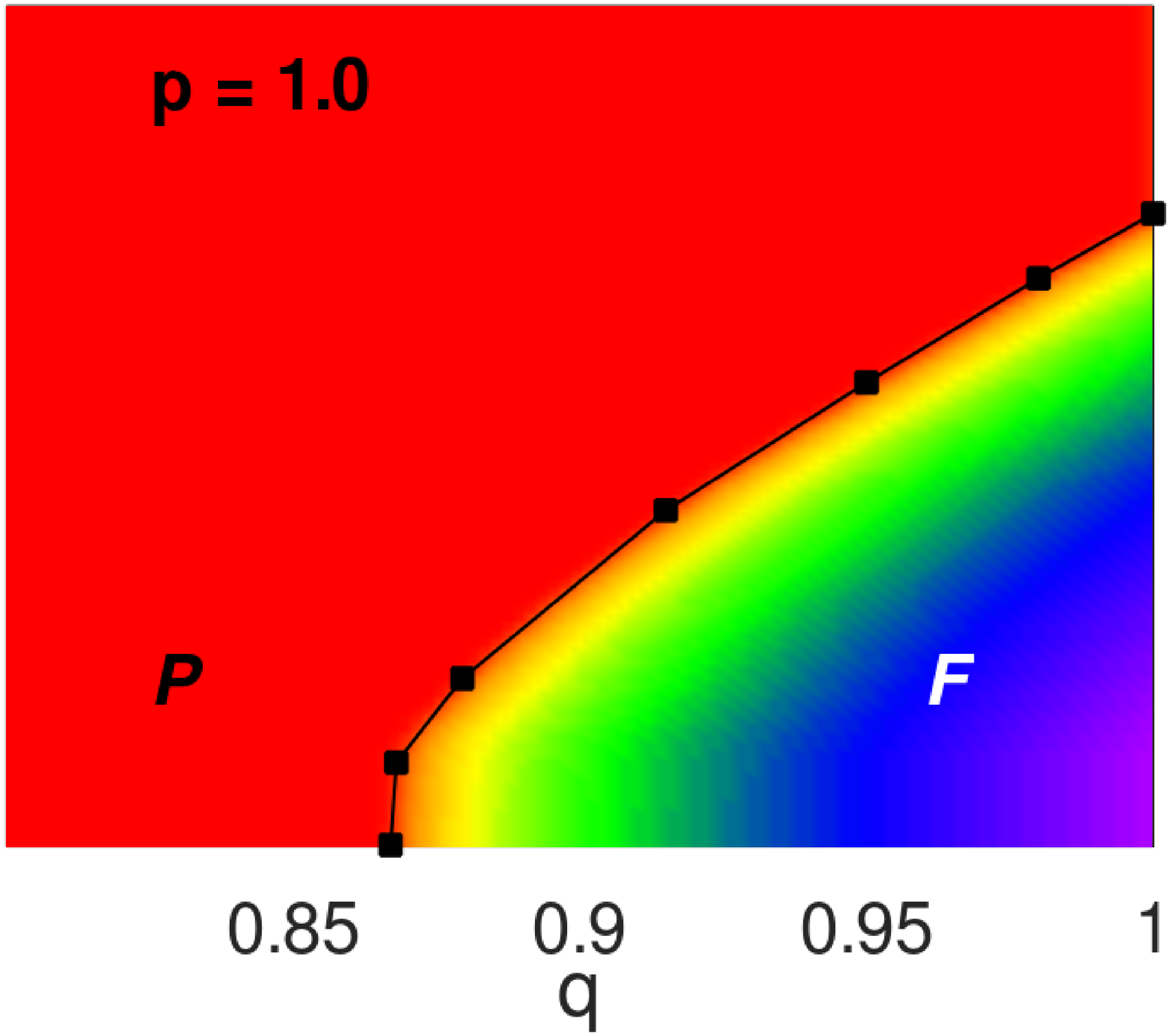}
\par\end{centering}
\caption{{\footnotesize{}Phase diagrams with different background colors for
the phase transitions between $AF-P$  and $P-F$  phases, and for
different values of $p$ as indicated in the figures. The black lines
are just a guide for the eyes in the second-order phase transitions
and the black square dots are critical points calculated. The color
map refers to the magnetization of the system, $\textrm{m}_{\textrm{L}}^{\textrm{AF}}$
and $\textrm{m}_{\textrm{L}}^{\textrm{F}}$, as indicated in the color
bars. \label{fig:5}}}
\end{figure}
\par\end{center}

\begin{center}
\begin{table}
\caption{{\footnotesize{}Critical competition probability $q_{c}$, based on
the extrapolating of the susceptibility discontinuity and in the $AF-P$
phase transition, for $T=1$\label{tab:1}}}

\centering{}%
\begin{tabular}{|c|c|c|c|c|}
\hline 
$p$ & $q_{c}^{\chi}$ & $\beta/\nu$ & $\gamma/\nu$ & $\nu$\tabularnewline
\hline 
\hline 
$0$ & $0.00149\pm0.0001$ & $0.16\pm0.05$ & $1.77\pm0.03$ & $1.04\pm0.09$\tabularnewline
\hline 
$0.25$ & $0.03966\pm0.002$ & $0.46\pm0.07$ & $1.07\pm0.06$ & $1.07\pm0.03$\tabularnewline
\hline 
$0.5$ & $0.07176\pm0.003$ & $0.45\pm0.04$ & $1.06\pm0.04$ & $0.96\pm0.09$\tabularnewline
\hline 
$0.75$ & $0.09062\pm0.002$ & $0.46\pm0.03$ & $1.04\pm0.02$ & $1.02\pm0.05$\tabularnewline
\hline 
$1.0$ & $0.09774\pm0.002$ & $0.48\pm0.02$ & $1.05\pm0.02$ & $1.06\pm0.08$\tabularnewline
\hline 
\end{tabular}
\end{table}
\par\end{center}

\begin{center}
\begin{table}
\caption{{\footnotesize{}Critical competition probability $q_{c}$, based on
the extrapolating of the susceptibility discontinuity and in the $F-P$
phase transition, for $T=1$.\label{tab:2}}}

\centering{}%
\begin{tabular}{|c|c|c|c|c|}
\hline 
$p$ & $q_{c}^{\chi}$ & $\beta/\nu$ & $\gamma/\nu$ & $\nu$\tabularnewline
\hline 
\hline 
$0$ & $0.964\pm0.002$ & $0.11\pm0.02$ & $1.67\pm0.09$ & $1.06\pm0.09$\tabularnewline
\hline 
$0.25$ & $0.917\pm0.002$ & $0.47\pm0.06$ & $1.01\pm0.06$ & $0.96\pm0.09$\tabularnewline
\hline 
$0.5$ & $0.895\pm0.001$ & $0.46\pm0.02$ & $1.04\pm0.02$ & $0.96\pm0.09$\tabularnewline
\hline 
$0.75$ & $0.883\pm0.002$ & $0.46\pm0.01$ & $1.06\pm0.01$ & $0.99\pm0.06$\tabularnewline
\hline 
$1.0$ & $0.880\pm0.001$ & $0.49\pm0.01$ & $1.0\pm0.02$ & $1.03\pm0.04$\tabularnewline
\hline 
\end{tabular}
\end{table}
\par\end{center}

With the critical point values, we built the phase diagram which shows
the regions on the plane of $T$ versus $q$, where the $F$, $P$,
and $AF$ phases are found. The phase diagrams are presented in Fig.
\ref{fig:5}, for different values of $p$, where we can see the greater
the probability of adding $J_{ik}$, the greater the region where
we find the ordered phases.

Now, in order to better understand the behavior of these phases (see
Fig. \ref{fig:5}), we can relate these ordered phases to dynamics
used in the competition. The $AF$ phase, observing the $q$-axis,
is found when $q\to0$ and the order parameter{\footnotesize{} $\textrm{m}_{\textrm{L}}^{\textrm{AF}}\rightarrow1$
}(see the figures on the left side in Fig. \ref{fig:5}), i.e., when
the two-spin flip dynamic prevails in the competition. This is because,
in the dynamic that simulates the system with an external energy flow
into it, the change in the spin states is only accepted if it increases
the energy of the system. Considering the Hamiltonian model, Eq. (\ref{eq:1}),
the state of the highest energy to which the dynamics lead the system
is the one where the spins are aligned antiparallel. The antiparallel
order also can be achieved through the A-SWN, because if we analyze
locally, the antiferromagnetic phase occurs when a central spin is
in the up (down) state, and its neighbors, to whom it is connected,
are in the down (up) state. Extending this analysis to the entire
network, an ordering of this type only occurs when we have well-defined
what are the central spins and what sites they can connect to, otherwise,
completely random long-range interactions can connect two distant
sites in the network that the highest local energy configuration of
one of these is unfavorable to the local antiparallel ordering of
the other site, thus, making it impossible to obtain the stationary
state with an $AF$ phase in the system. In this context, the $F$
phase is found in the limit that $q\to1$ and the order parameter{\footnotesize{}
$\textrm{m}_{\textrm{L}}^{\textrm{F}}\rightarrow1$ }(see the figures
of the right side in Fig. \ref{fig:5}), i.e., when the one-spin flip
dynamic prevails. This dynamic is responsible to simulate the system
in contact with the heat bath at temperature $T$, and favors the
lowest energy state through the thermal equilibrium, in which all
spins have the same state following the Hamiltonian system, so, if
we wanted to, we could treat them without the sublattices in the A-SWN
regime. On the other hand, when none of the dynamics prevails, i.e.,
between the extremes of the probability $q$-value, no one of the
expected order phase types is found in the system. Thus, we have most
of the values of $q$, the $P$ phase in the system is found, where
both{\footnotesize{} $\textrm{m}_{\textrm{L}}^{\textrm{AF}}\rightarrow0$
}and{\footnotesize{} $\textrm{m}_{\textrm{L}}^{\textrm{F}}\rightarrow0$}.
Another important observation is that the phase diagram topology changes
when $p$ increases, but the phases do not disappear.
\begin{center}
\begin{table*}
\caption{{\footnotesize{}Critical competition probability $q_{c}$, based on
the crossing of the fourth-order Binder cumulant curves and in the
$AF-P$ phase transition, for $T=1$. The effective dimension is given
by hyperscaling relation $d_{\textrm{eff}}=2\beta/\nu+\gamma/\nu$.
\label{tab:3}}}

\centering{}%
\begin{tabular}{|c|c|c|c|c|c|c|}
\hline 
$p$ & $q_{c}^{U}$ & $\beta$ & $\nu\left(\textrm{\ensuremath{\textrm{m}^{\textrm{AF}}}}\right)$ & $\gamma$ & $\nu\left(\chi^{\textrm{AF}}\right)$ & $d_{\textrm{eff}}$\tabularnewline
\hline 
\hline 
$0$ & $0.00145\pm0.00004$ & $0.125\pm0.03$ & $1.0\pm0.05$ & $1.75\pm0.04$ & $1.0\pm0.05$ & $2.00\pm0.21$\tabularnewline
\hline 
$0.25$ & $0.0390\pm0.0009$ & $0.46\pm0.06$ & $0.96\pm0.06$ & $1.06\pm0.05$ & $1.0\pm0.05$ & $2.02\pm0.29$\tabularnewline
\hline 
$0.5$ & $0.0715\pm0.0006$ & $0.44\pm0.05$ & $1.0\pm0.05$ & $1.05\pm0.03$ & $1.0\pm0.04$ & $1.93\pm0.24$\tabularnewline
\hline 
$0.75$ & $0.0921\pm0.0007$ & $0.48\pm0.04$ & $0.98\pm0.06$ & $1.05\pm0.06$ & $1.0\pm0.04$ & $2.03\pm0.31$\tabularnewline
\hline 
$1.0$ & $0.0978\pm0.0004$ & $0.48\pm0.03$ & $1.02\pm0.06$ & $1.06\pm0.06$ & $0.98\pm0.04$ & $2.02\pm0.23$\tabularnewline
\hline 
\end{tabular}
\end{table*}
\par\end{center}

\begin{center}
\begin{table*}
\caption{{\footnotesize{}Critical competition probability $q_{c}$, based on
the crossing of the fourth-order Binder cumulant curves and in the
$F-P$ phase transition, for $T=1$. The effective dimension is given
by hyperscaling relation $d_{\textrm{eff}}=2\beta/\nu+\gamma/\nu$.
\label{tab:4}}}

\centering{}%
\begin{tabular}{|c|c|c|c|c|c|c|}
\hline 
$p$ & $q_{c}^{U}$ & $\beta$ & $\nu\left(\textrm{\ensuremath{\textrm{m}^{\textrm{F}}}}\right)$ & $\gamma$ & $\nu\left(\chi^{\textrm{F}}\right)$ & $d_{\textrm{eff}}$\tabularnewline
\hline 
\hline 
$0$ & $0.965\pm0.0004$ & $0.125\pm0.02$ & $1.0\pm0.05$ & $1.70\pm0.07$ & $1.01\pm0.03$ & $1.95\pm0.20$\tabularnewline
\hline 
$0.25$ & $0.917\pm0.0009$ & $0.45\pm0.05$ & $1.0\pm0.05$ & $1.05\pm0.05$ & $1.0\pm0.04$ & $1.95\pm0.26$\tabularnewline
\hline 
$0.5$ & $0.895\pm0.0008$ & $0.46\pm0.04$ & $0.95\pm0.07$ & $1.05\pm0.04$ & $1.0\pm0.04$ & $2.02\pm0.12$\tabularnewline
\hline 
$0.75$ & $0.883\pm0.0005$ & $0.47\pm0.03$ & $1.0\pm0.06$ & $1.04\pm0.06$ & $1.0\pm0.05$ & $1.98\pm0.23$\tabularnewline
\hline 
$1.0$ & $0.879\pm0.0006$ & $0.49\pm0.04$ & $1.0\pm0.07$ & $1.02\pm0.05$ & $0.99\pm0.06$ & $2.00\pm0.26$\tabularnewline
\hline 
\end{tabular}
\end{table*}
\par\end{center}

\begin{center}
\begin{figure}
\begin{centering}
\includegraphics[scale=0.32]{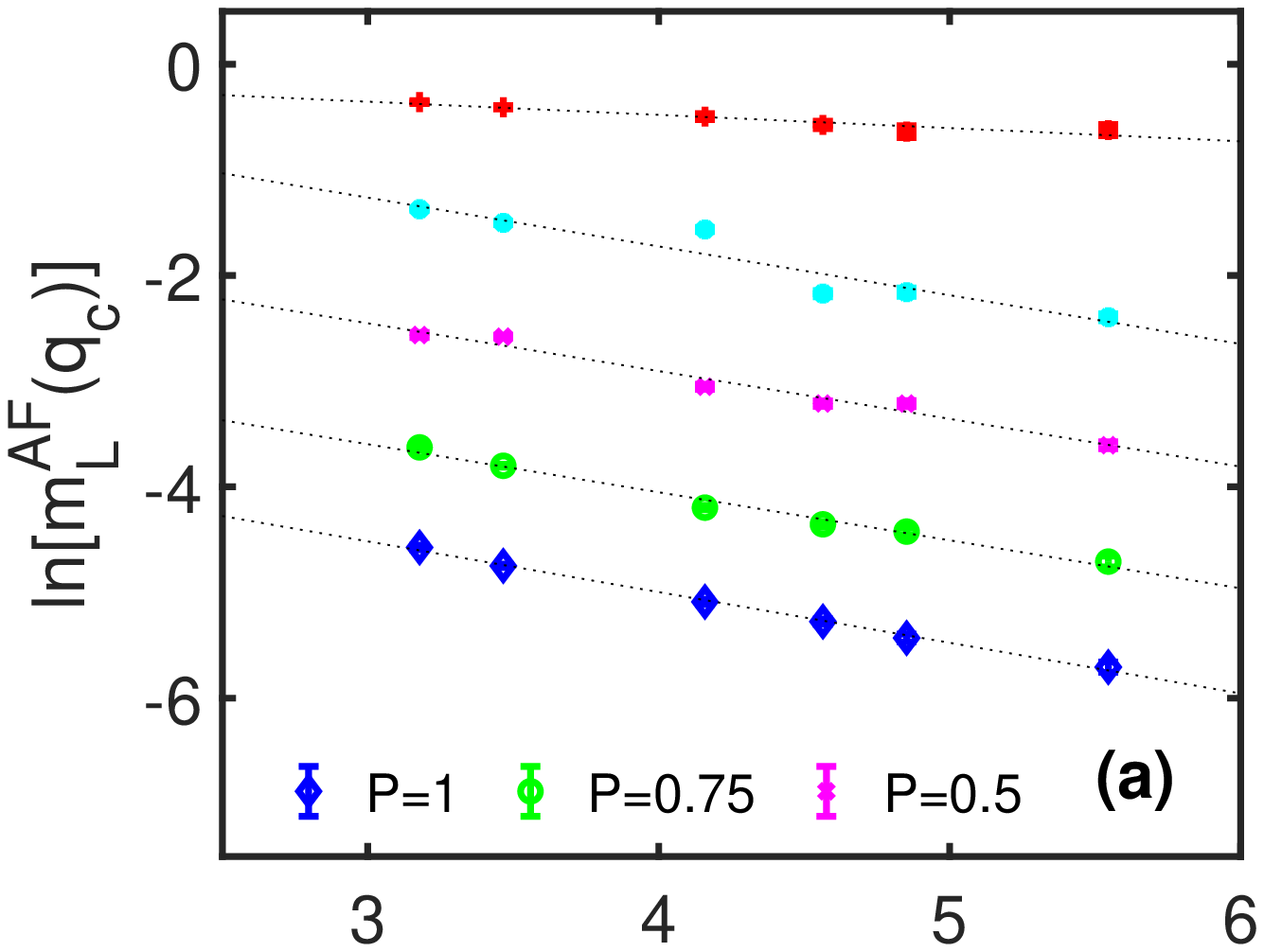}\includegraphics[scale=0.32]{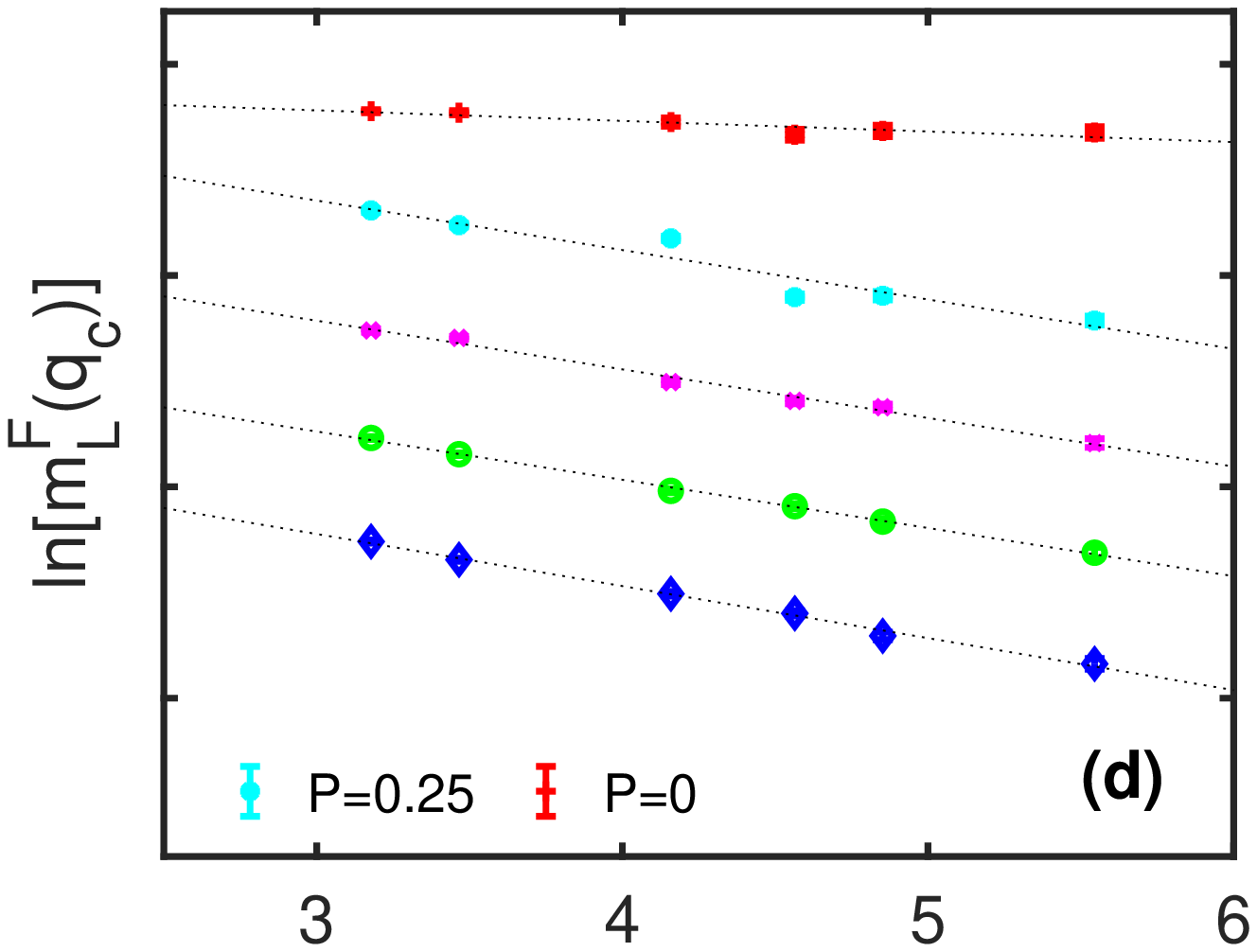}
\par\end{centering}
\begin{centering}
\includegraphics[scale=0.32]{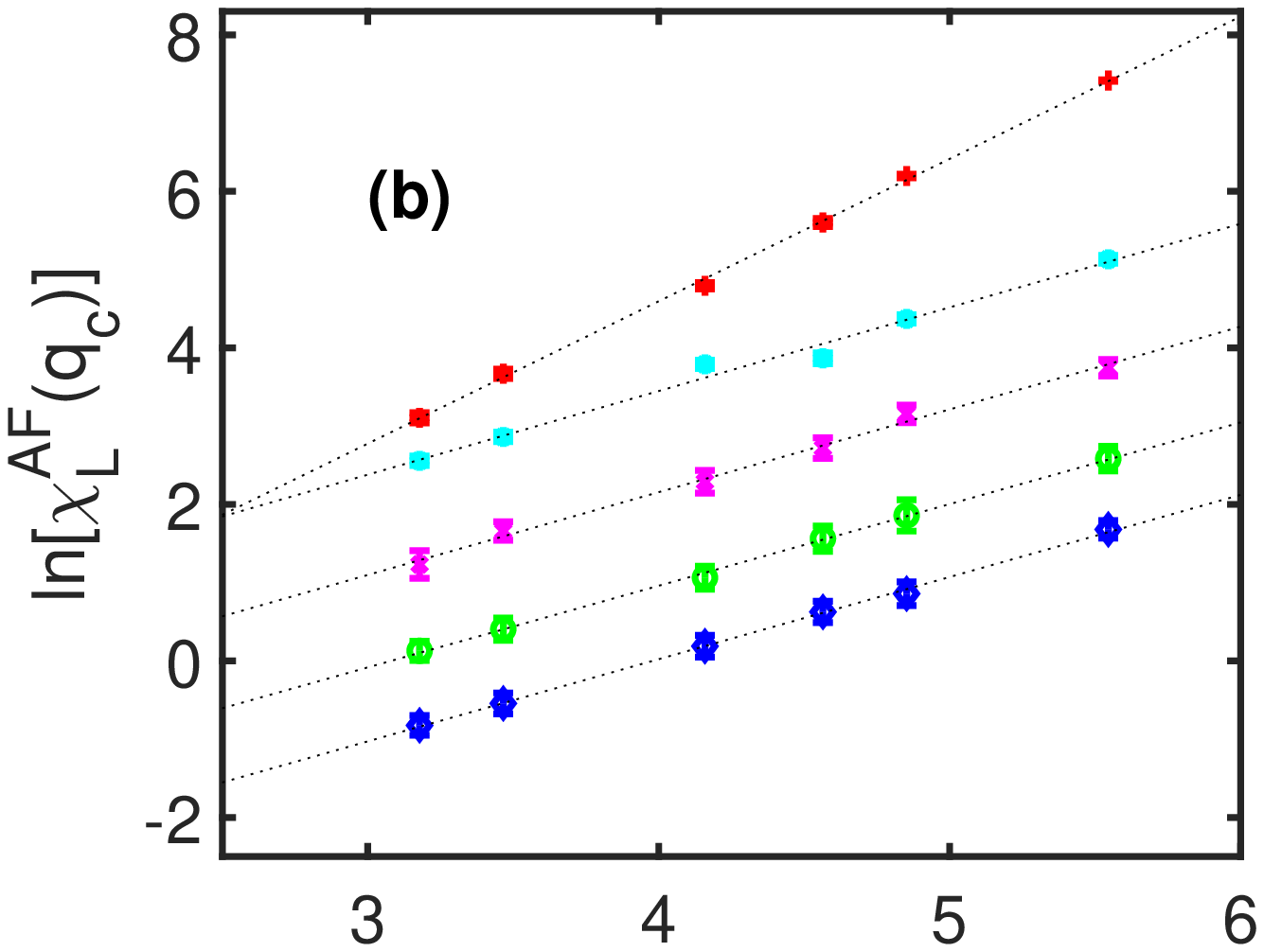}\includegraphics[scale=0.32]{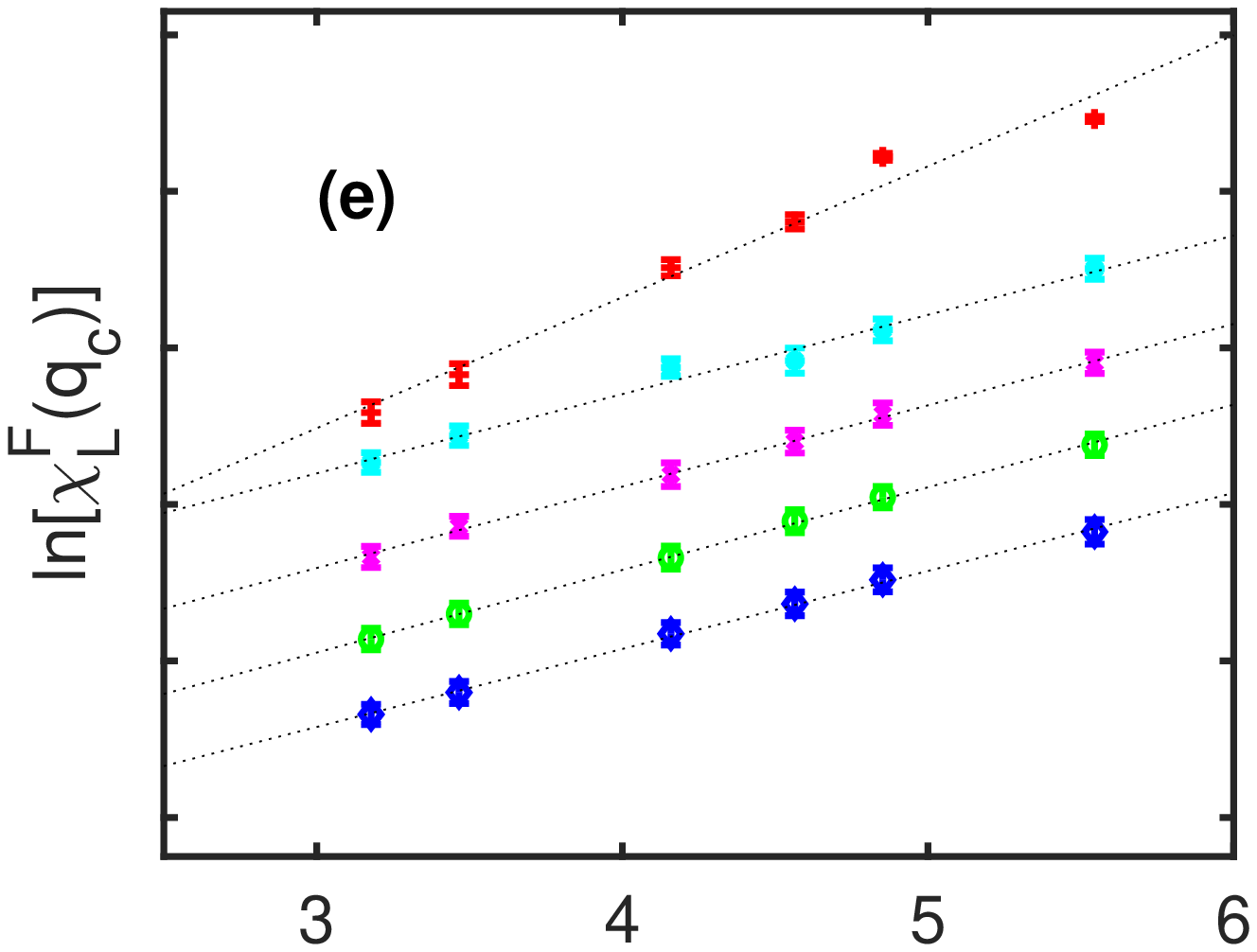}
\par\end{centering}
\begin{centering}
\includegraphics[scale=0.32]{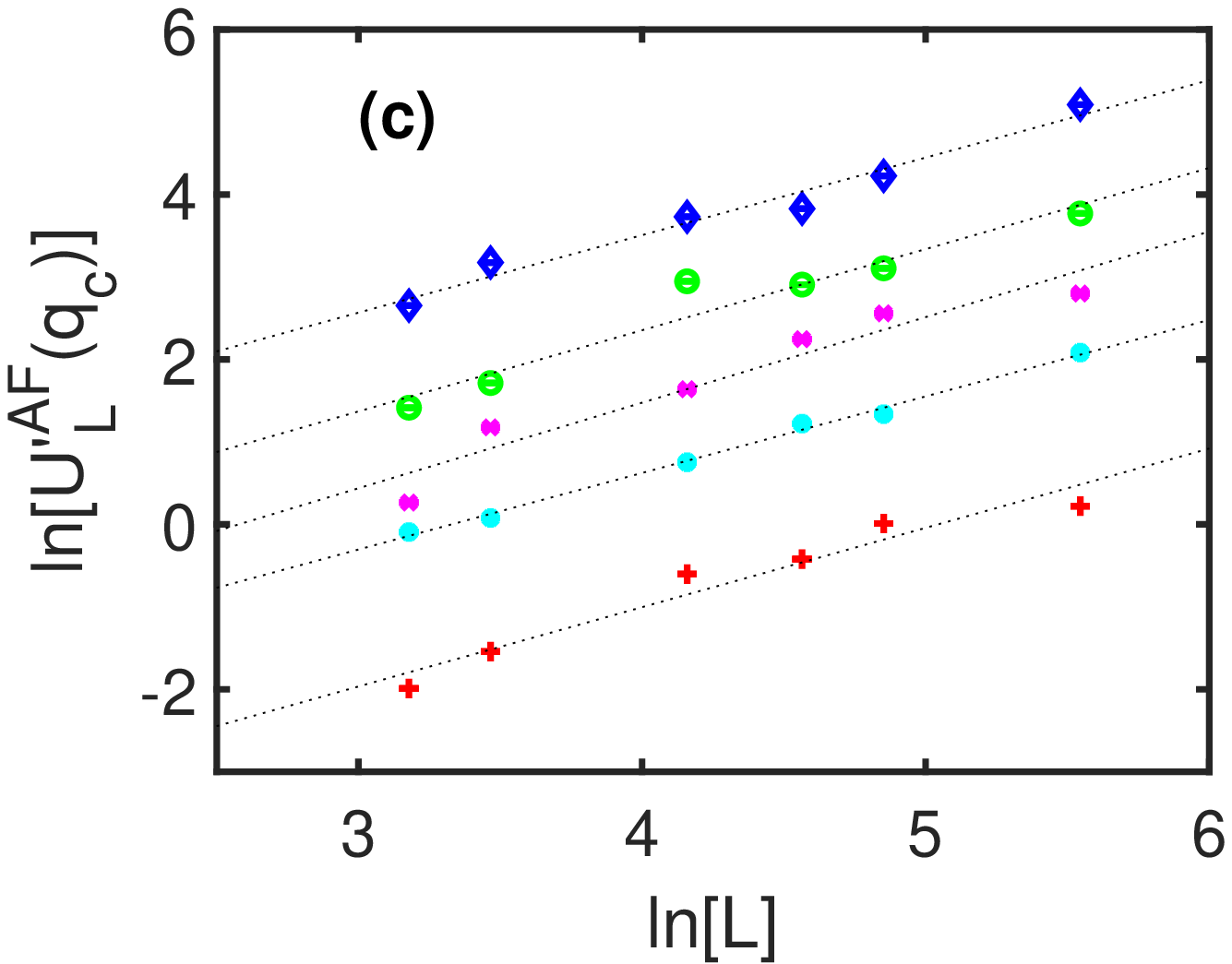}\includegraphics[scale=0.32]{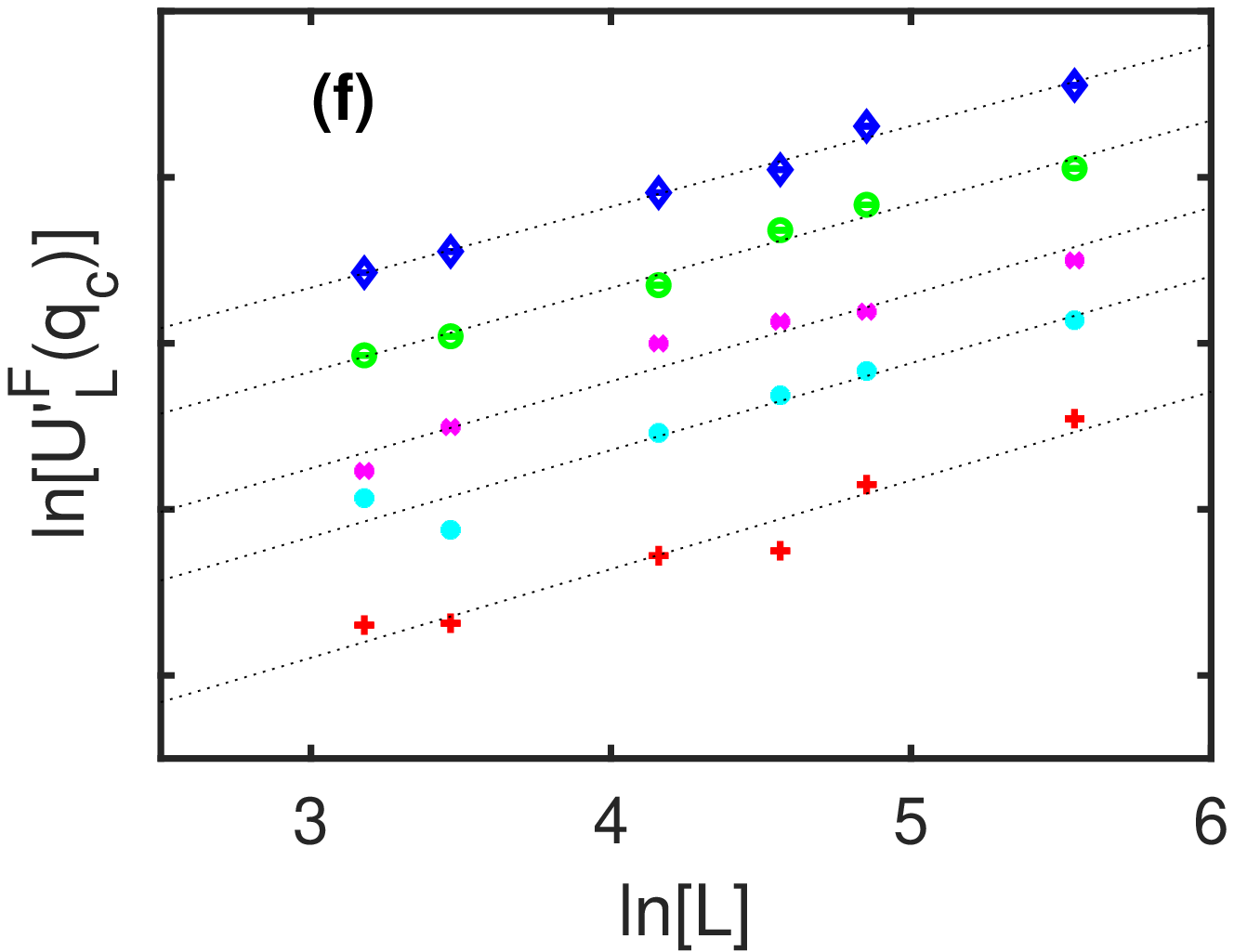}
\par\end{centering}
\caption{{\footnotesize{}(a) and (d) contains the best fit of the staggered
$\textrm{m}_{\textrm{L}}^{\textrm{AF}}$ and total $\textrm{m}_{\textrm{L}}^{\textrm{F}}$
magnetization curves in the critical point as a function of the different
values of $L$ in the log-log plot, respectively, for the $F$ and
$AF$ phases, in which the slopes we obtained the ratio $\beta/\nu$
between the critical exponents. (b) and (e) in the $F$ and $AF$
phases, we have the best fit of the staggered $\chi_{\textrm{L}}^{\textrm{AF}}$
and total $\textrm{\ensuremath{\chi}}_{\textrm{L}}^{\textrm{F}}$
susceptibility curves at the critical point as a function of $L$,
and the slopes have given us the ratio $\gamma/\nu$. The critical
exponent $\nu$, we have obtained by the slope of the best fit of
the Binder cumulant derivative $\textrm{U'}_{\textrm{L}}^{\textrm{AF}}$
and $\textrm{U'}_{\textrm{L}}^{\textrm{F}}$ at the critical point,
as represented in (c) and (f) for the $F$ and $AF$ phases, respectively.
The critical exponents obtained by this method for the $AF$ phase
can be seen in Table \ref{tab:1}, and in Table \ref{tab:2} for the
$F$ phase. \label{fig:6}}}
\end{figure}
\par\end{center}

After the presentation of the phase diagrams by exploiting the thermal
variations of the order parameters, the Binder cumulant and the magnetic
susceptibility, we can now study the critical behavior of these quantities
in the vicinity of the phase transitions using the FSS method to evaluate
some critical exponents of the model. Therefore, to obtain the critical
exponents, we also used two methods, both referring to the FSS method,
using the scale relations of Eqs. (\ref{eq:11}), (\ref{eq:12}),
and (\ref{eq:14}). One of the methods refers to the value of the
thermodynamic quantities at the critical point, in which when we make
a log-log plot of the value of these quantities as a function of $L$.
Using the scale relations, we obtain ratios between the critical exponents
through the slope of the line of best fit of those points. In Fig.
\ref{fig:6}, the behavior of thermodynamic quantities near the critical
point can be seen as a function of $L$ in the log-log plot. In Figs.
\ref{fig:6}(a) and \ref{fig:6}(d) we were able to find the ratio
$-\beta/\nu$ in the $AF-P$ and $F-P$ phase transitions, respectively,
using the scaling relation of the Eq. (\ref{eq:11}), through the
slope in the linear fit of the points for each selected value of $p$,
as indicated in the figures. The same can be done using the scaling
relation of the Eq. (\ref{eq:12}), however, the critical exponent
ratio is $\gamma/\nu$ and obtained by the slope of the linear fits
of Figs. \ref{fig:6}(b) and \ref{fig:6}(e), for the different values
of $p$ and in the $AF-P$ and $F-P$ phase transitions, respectively.
Finally, the ratios between the exponents obtained previously, it
is useful to use the scaling relation of Eq. (\ref{eq:14}), which
we have used the data of the Binder cumulant derivative to obtain
information related to the critical exponent of correlation length,
$\nu$. Here, they were obtained from the linear fit of the curves
of Figs. \ref{fig:6}(c) and \ref{fig:6}(f) for the different $p$
values and $AF-P$ and $F-P$ phase transitions, respectively. It
is worth noting that as our interest is in the slope of the log-log
plot, we changed the linear coefficients of the straight lines to
separate the lines and thus making it easier for the reader to see
the fits. All the ratios between the values of the critical exponents
obtained by the log-log plot of the scaling relations can be seen
in Table \ref{tab:1} for the $AF-P$ phase transitions, and in Table
\ref{tab:2} for the $F-P$ phase transitions. 
\begin{center}
\begin{figure}
\begin{centering}
\includegraphics[clip,scale=0.31]{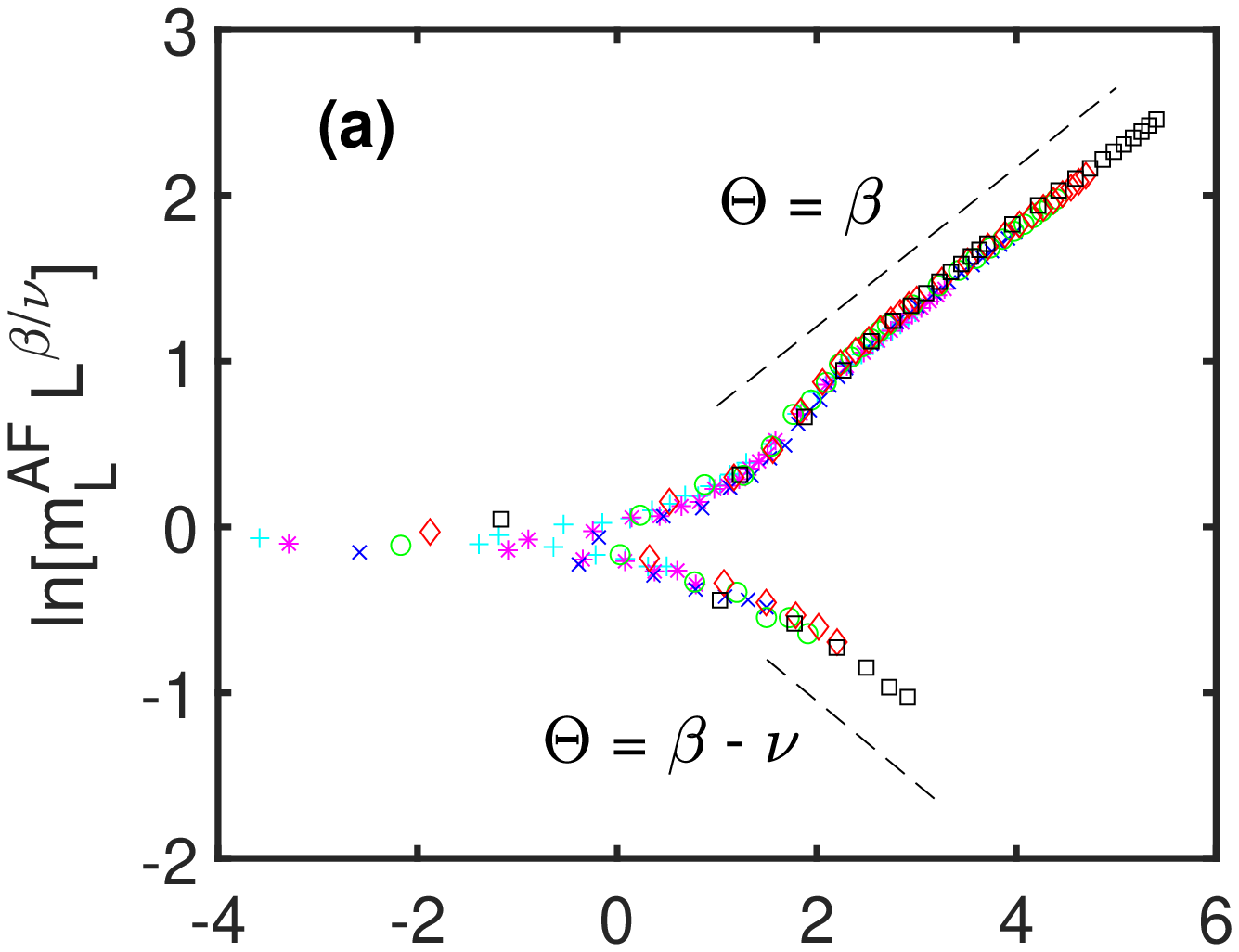}\includegraphics[clip,scale=0.31]{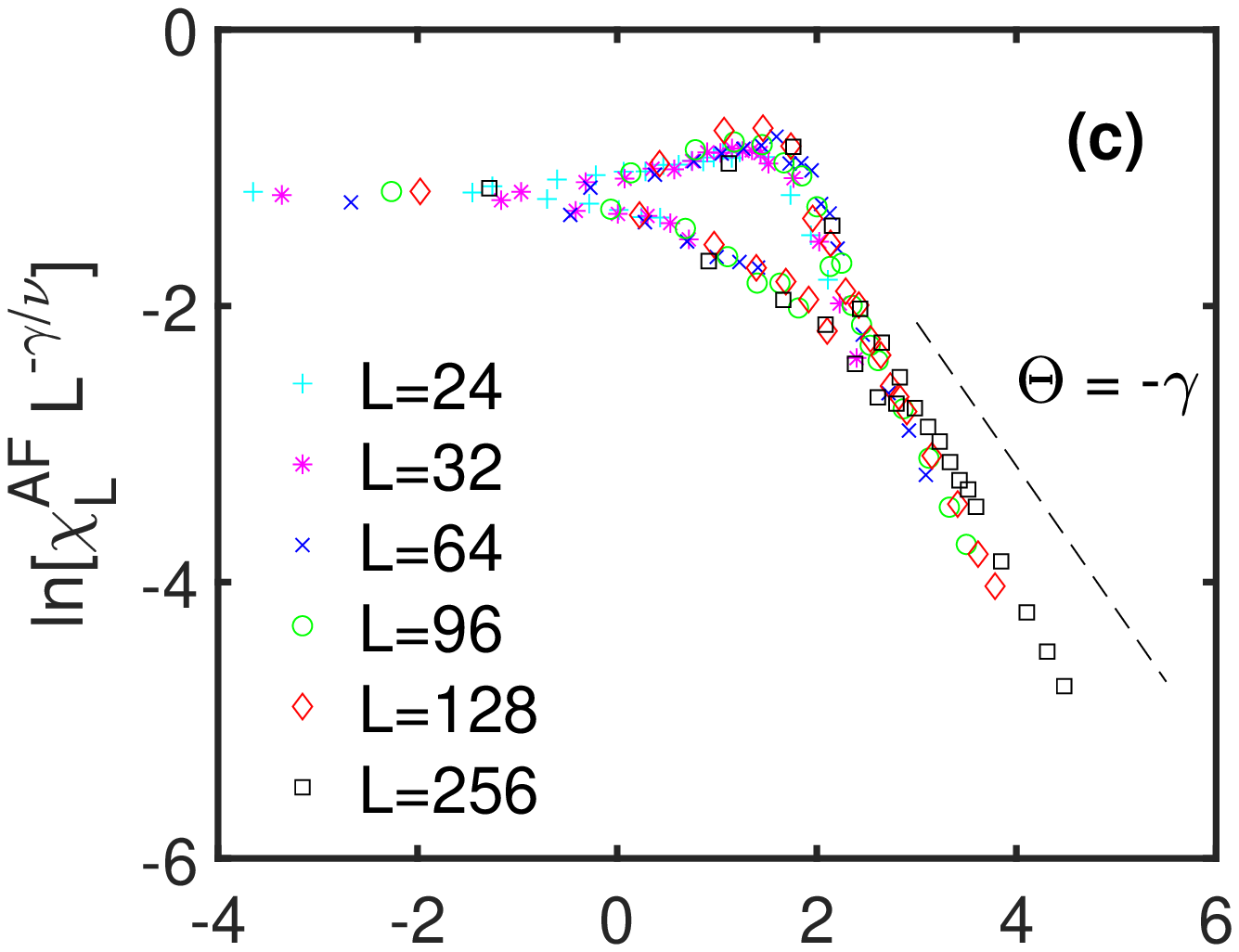}
\par\end{centering}
\begin{centering}
\includegraphics[clip,scale=0.31]{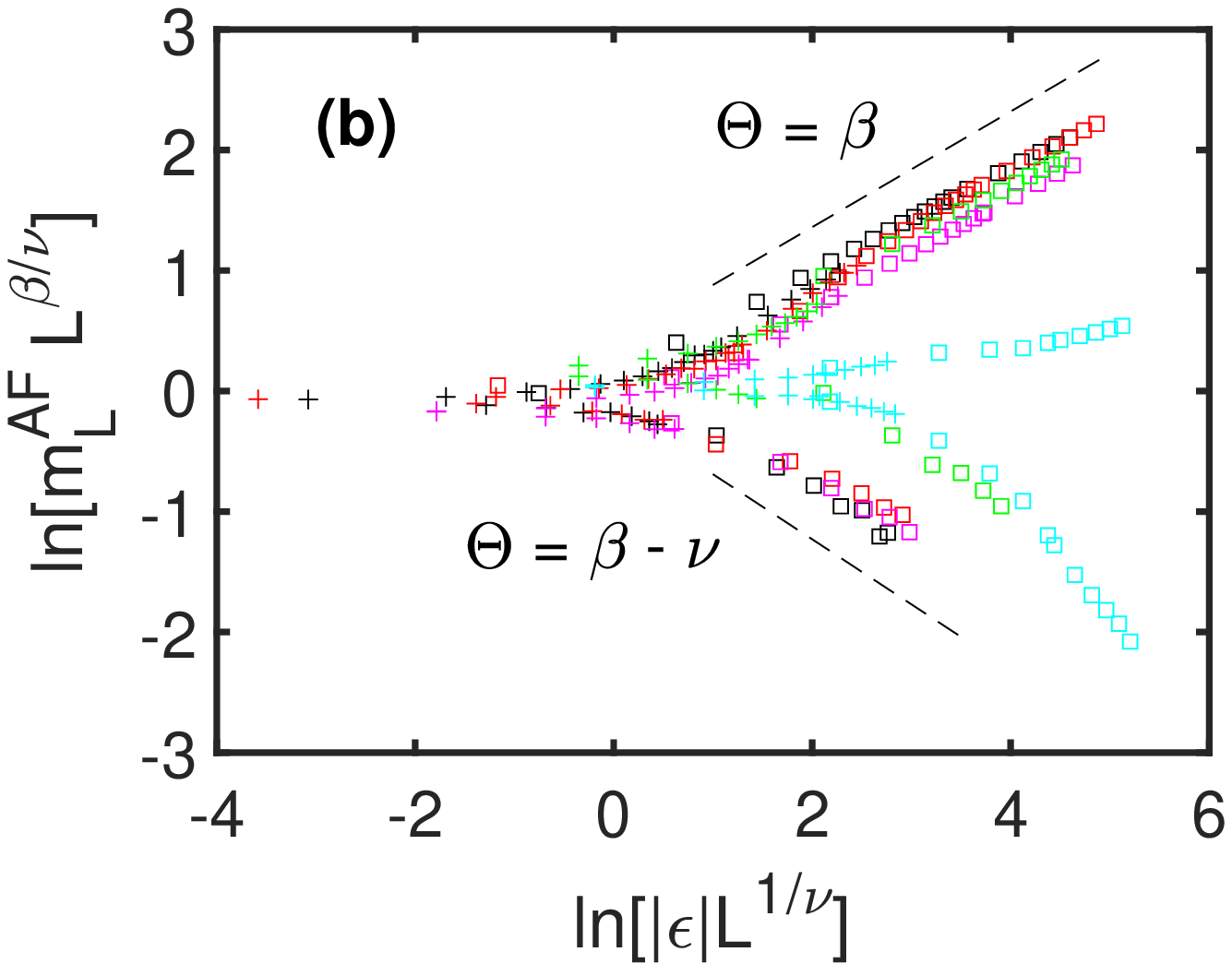}\includegraphics[clip,scale=0.31]{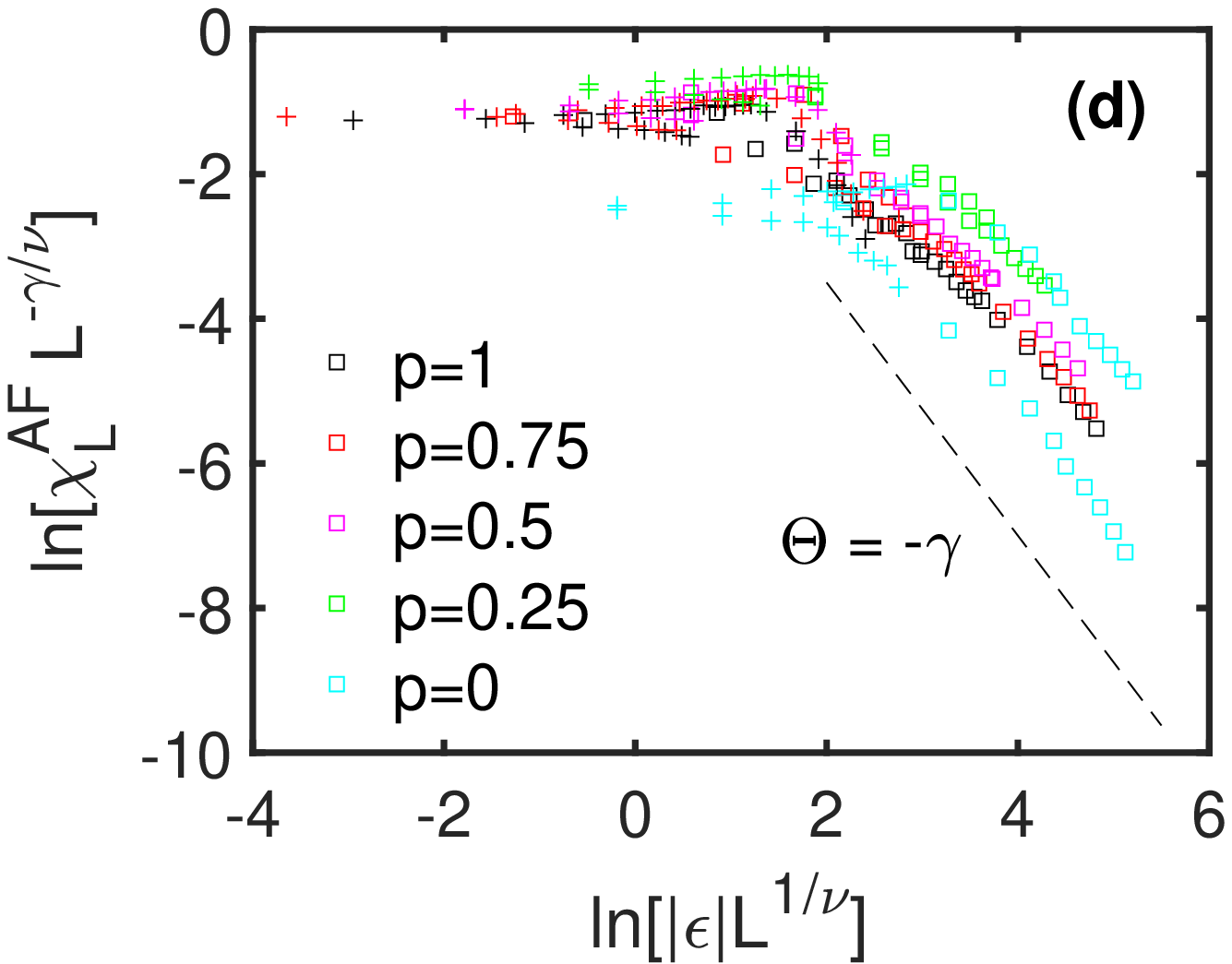}
\par\end{centering}
\caption{{\footnotesize{}Data collapse by FSS analysis for (a) staggered magnetization
$\textrm{m}_{\textrm{L}}^{\textrm{AF}}$ and (c) susceptibility $\textrm{\ensuremath{\chi}}_{\textrm{L}}^{\textrm{AF}}$
for different values of $L$ as indicated in the figure, and fixed
$p=0.75$. In (b) and (d) we have the data collapse for $\textrm{m}_{\textrm{L}}^{\textrm{AF}}$
and $\textrm{\ensuremath{\chi}}_{\textrm{L}}^{\textrm{AF}}$, respectively,
but for all values of $p$, in which the critical exponents were obtained
using the best data collapse for all lattice sizes, and here we only
display the lattice sizes $L=24\:(+)$ and $L=256\:(\square)$. The
$\varepsilon$ parameter is set to $\varepsilon=(q-q_{c})/q_{c}$.
The dashed lines represent the asymptotic behavior of the scale functions.
The values of the critical exponents $\beta$, $\gamma$, $\nu\left(\textrm{\ensuremath{\textrm{m}^{\textrm{AF}}}}\right)$,
and $\nu\left(\chi^{\textrm{AF}}\right)$ of the best data collapse
can be seen in Table \ref{tab:3}. \label{fig:7}}}
\end{figure}
\par\end{center}

Another method used to obtain the critical exponents is through the
scaling functions in Eqs. (\ref{eq:11}), (\ref{eq:12}) and (\ref{eq:14}),
in the around of the critical point. For this, we isolate the scale
function and plot it in a log-log plot through the curves of $\textrm{\ensuremath{\textrm{m}^{\textrm{F}}}}L^{\beta/\nu}$
and $\textrm{\ensuremath{\textrm{m}^{\textrm{AF}}}}L^{\beta/\nu}$
as a function of $\left|\varepsilon\right|L^{1/\nu}$, resulting in
a single curve for all lattice sizes $L$ if we have the correct critical
exponents and critical points adjusted in the scaling relations. In
this method, the data collapse can also be obtained in a plot that
does not have the axes on the logarithmic scale, but the asymptotic
behavior that relates to the critical exponents are not present. We
can obtain the critical exponents because the data collapse in the
vicinity of the critical point, depends on the correct critical exponents
of the system to occur, in this way, we adjust them to obtain the
best data collapse  in the criticality, and consequently, the exponents
involved in this data collapses are the critical exponents of the
system. All values of the critical exponents obtained by data collapse
of the scaling relations can be seen in Table \ref{tab:3} for the
$AF-P$  phase transitions $\left(\beta,\nu\left(\textrm{\ensuremath{\textrm{m}^{\textrm{AF}}}}\right),\nu\left(\chi^{\textrm{AF}}\right),\gamma\right)$,
and in Table \ref{tab:4} for the $F-P$ phase transitions $\left(\beta,\nu\left(\textrm{\ensuremath{\textrm{m}^{\textrm{F}}}}\right),\nu\left(\chi^{\textrm{F}}\right),\gamma\right)$. 

In Fig. \ref{fig:7}, we have shown the data collapse for the scaling
functions of magnetization, Fig. \ref{fig:7}(a), and for magnetic
susceptibility, Fig. \ref{fig:7}(c), for the $AF-P$ phase transition,
with $p=0.75$, which was the best data collapse obtained. In this
phase transition, we also plotted for all values of $0\leq p\leq1.0$,
as can be seen in Fig. \ref{fig:7}(b) and \ref{fig:7}(d), the scaling
function of magnetization and magnetic susceptibility, respectively,
showing the best data collapse for the selected $p$ values, but displaying
only the lattice sizes $L=24\:(+)$ and $L=256\:(\square)$ for the
best differentiation between the collapsed curves.
\begin{center}
\begin{figure}
\begin{centering}
\includegraphics[clip,scale=0.31]{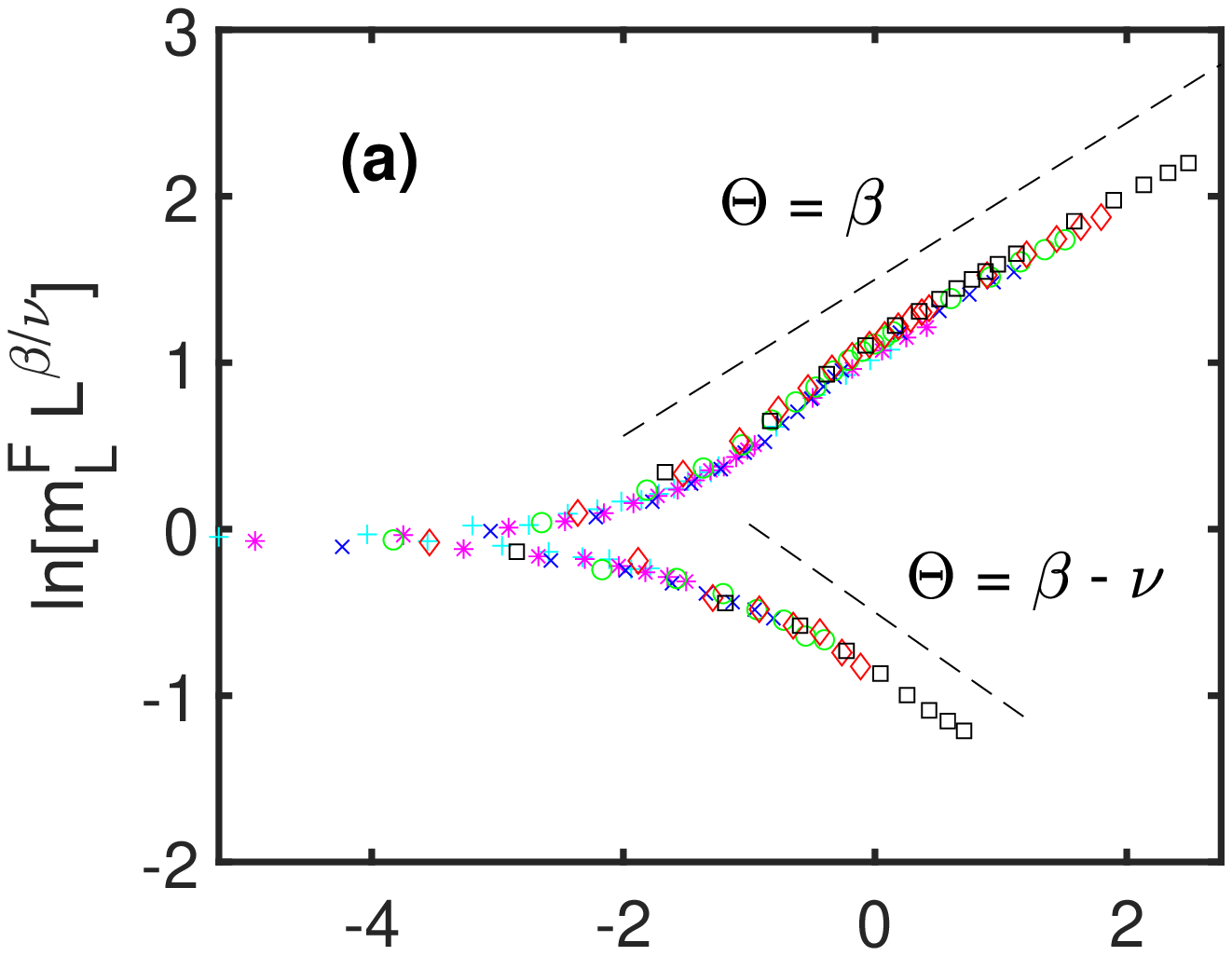}\includegraphics[clip,scale=0.31]{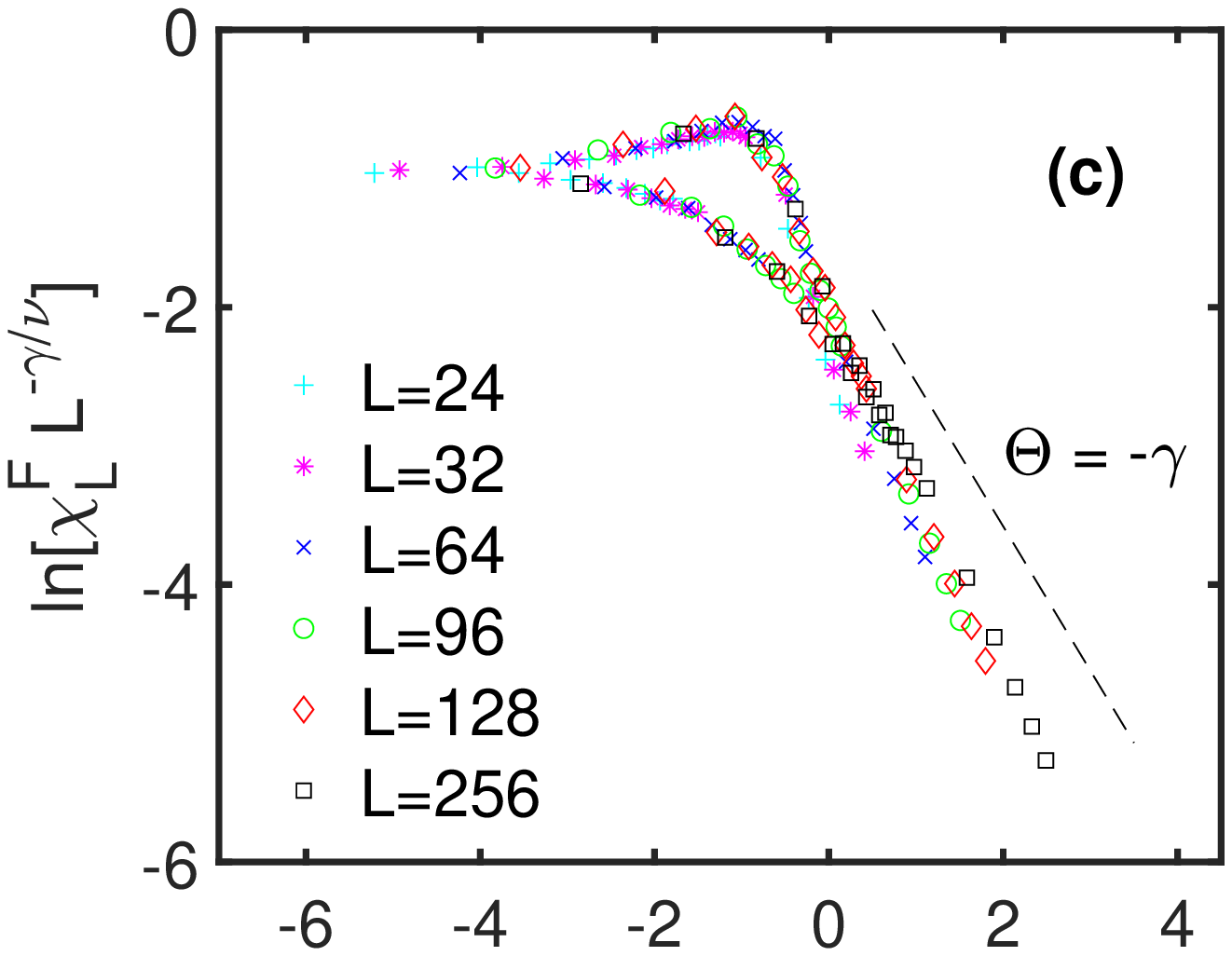}
\par\end{centering}
\begin{centering}
\includegraphics[clip,scale=0.31]{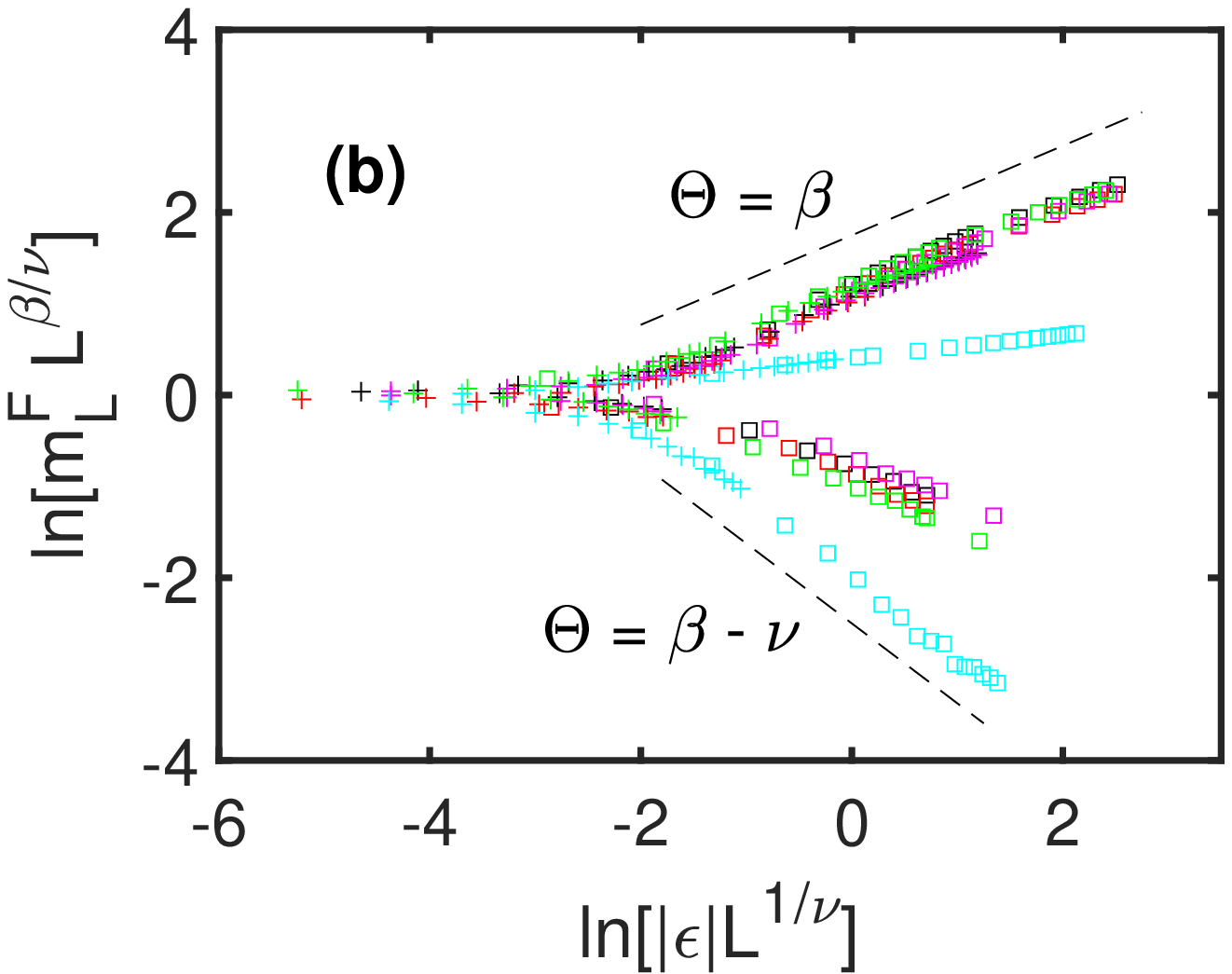}\includegraphics[clip,scale=0.31]{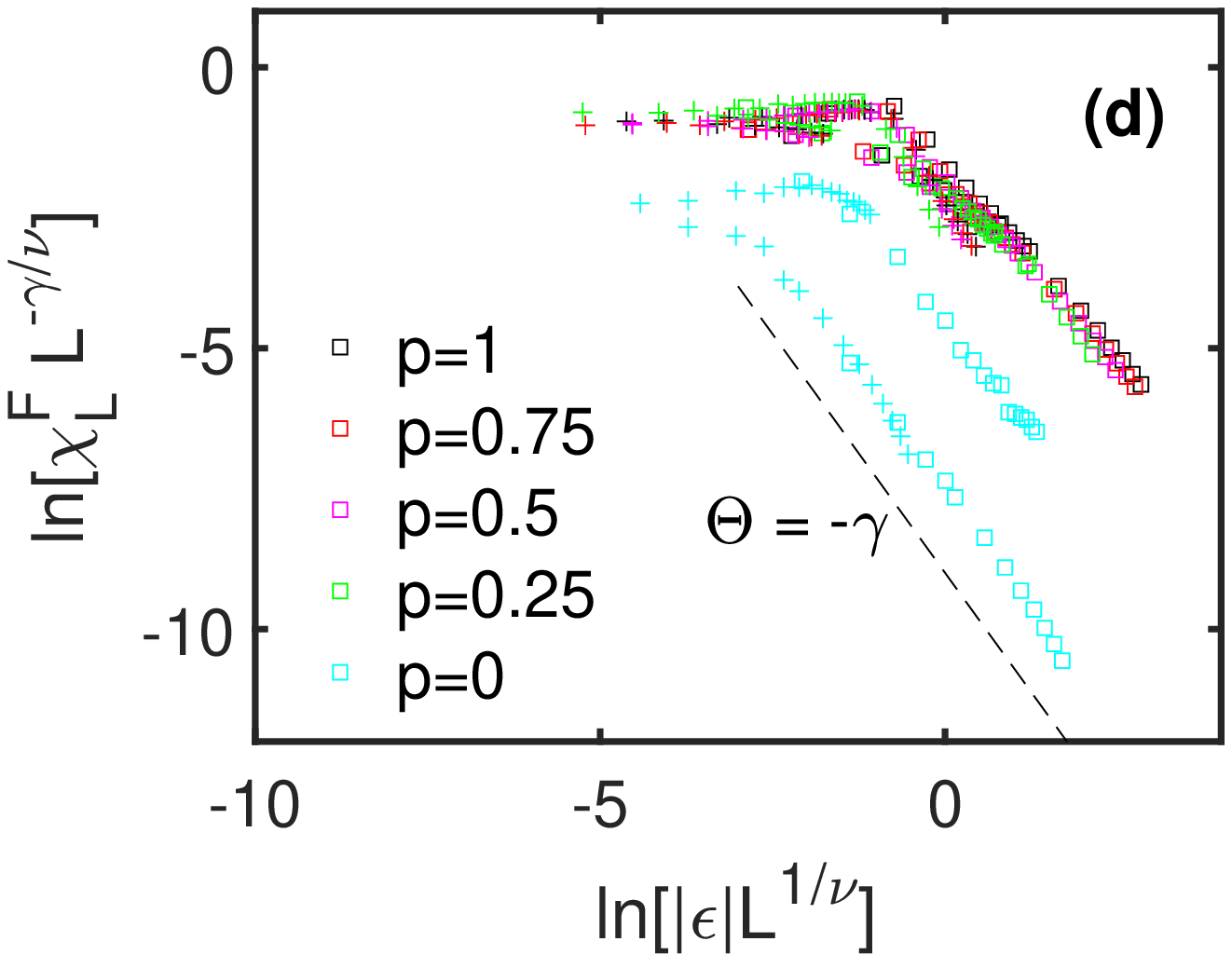}
\par\end{centering}
\caption{{\footnotesize{}Data collapse by FSS analysis for (a) total magnetization
$\textrm{m}_{\textrm{L}}^{\textrm{F}}$ and (c) susceptibility $\textrm{\ensuremath{\chi}}_{\textrm{L}}^{\textrm{F}}$
for different values of $L$ as indicated in the figure, and fixed
$p=0.75$. In (b) and (d) we have the data collapse for $\textrm{m}_{\textrm{L}}^{\textrm{F}}$
and $\textrm{\ensuremath{\chi}}_{\textrm{L}}^{\textrm{F}}$, respectively,
but for all values of $p$, in which the critical exponents were obtained
using the best data collapse  for all lattice sizes, and here we only
display the lattice sizes $L=24\:(+)$ and $L=256\:(\square)$. The
$\varepsilon$ parameter is set to $\varepsilon=(q-q_{c})/q_{c}$.
The dashed lines represent the asymptotic behavior of the scale functions.
The values of the critical exponents $\beta$, $\gamma$, $\nu\left(\textrm{\ensuremath{\textrm{m}^{\textrm{F}}}}\right)$,
and $\nu\left(\chi^{\textrm{F}}\right)$ of the best data collapse
can be seen in Table \ref{tab:4}. \label{fig:8}}}
\end{figure}
\par\end{center}

However, in the $F-P$ phase transition, the figures that present
the best data collapse, for $p=0.75$, are Fig. \ref{fig:8}(a) for
the magnetization scaling function, and Fig. \ref{fig:8}(c) for the
magnetic susceptibility scaling function. We also have calculated
for all values of $0\leq p\leq1.0$, but also only displaying here
two lattice sizes, $L=24\:(+)$ and $L=256\:(\square)$, and they
can be found in Figs. \ref{fig:8}(b) and \ref{fig:8}(d) for the
magnetization and the magnetic susceptibility scaling function, respectively.
The critical exponents obtained by this method, data collapse, can
be found in Table \ref{tab:3} for $AF-P$ phase transitions, and
in Table \ref{tab:4} for phase transitions from $F$ to $P$.

In both methods, we obtained very approximate values for the critical
exponents for the selected $p$ values. The best results shown here
are based on data collapse, this is due to the fact that for $p=0$
the values of the critical exponents are closer to the values of the
Ising model on the regular square lattice, which they are known by
exact solution and MC simulation, $\beta=1/8$, $\gamma=7/4$, and
$\nu=1$. On the other hand, when we increase the additive probability
$p$, we also increase the number of $J_{ik}$ added to the system,
thus, it is convenient to use the scaling relations of systems that
can have mean-field critical behavior, by the prediction that we have
a system above the Ising model critical dimension, $d=4$, in the
A-SWN regime. To do that, it is enough in the scaling relations, Eqs.
(\ref{eq:11}), (\ref{eq:12}), and (\ref{eq:14}), to substitute
the linear length $L$ of the lattice by the total number of spins
in the system, $L^{2}=N$. By doing this, as predicted, are obtained
approximately the mean-field critical exponents $\beta=1/2$, $\gamma=1$
and $\nu=1/2$. The behavior of both the critical exponents for $p=0$
and the A-SWN regime $(0<p\leq1)$ were represented in Fig. \ref{fig:9}(a)
for the $AF-P$ phase transition and in Fig. \ref{fig:9}(b) for the
$F-P$ phase transition.
\begin{center}
\begin{figure}
\begin{centering}
\includegraphics[scale=0.5]{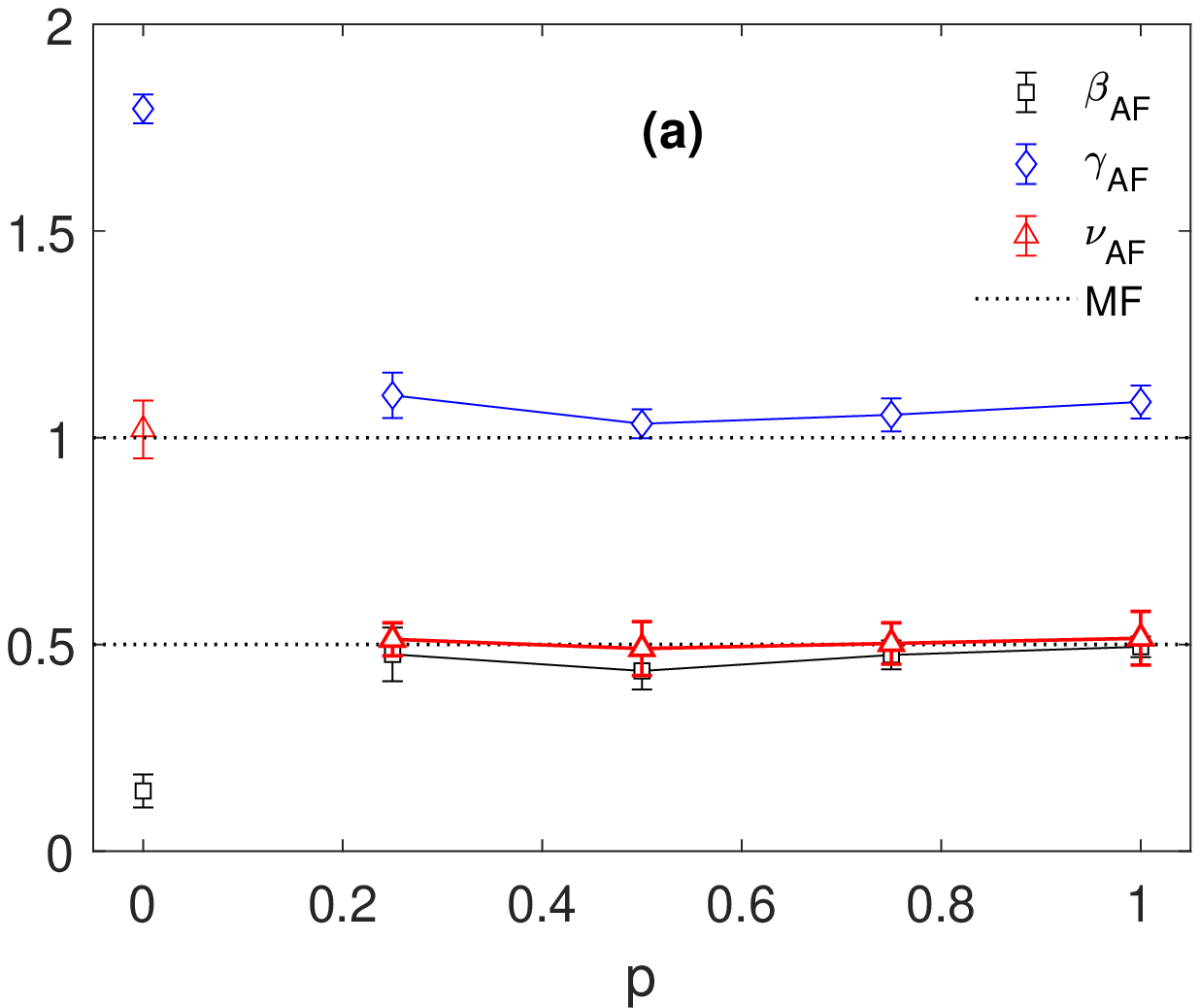}
\par\end{centering}
\begin{centering}
\vspace{0.1cm}
\par\end{centering}
\begin{centering}
\includegraphics[scale=0.5]{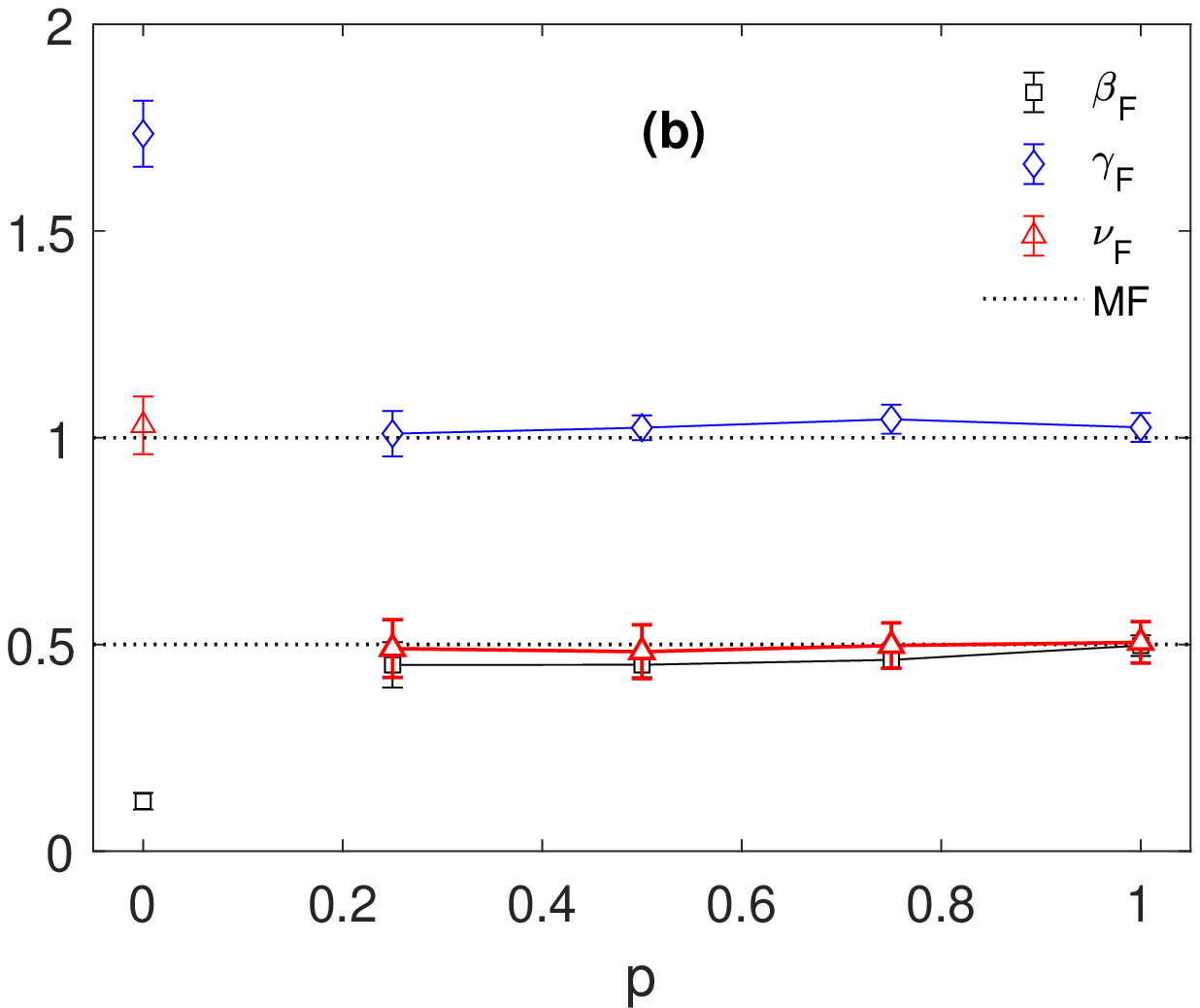}
\par\end{centering}
\caption{{\footnotesize{}(a) Representation of the critical exponents presented
in Tables \ref{tab:3} and \ref{tab:1} for the critical behavior
of the system in the $AF-P$  phase transition, taking into account
the mean-field scale relationships for the A-SWN regime $(0<p\le1)$.
(b) Representation of the critical exponents presented in Tables \ref{tab:4}
and \ref{tab:2} for the system in the $F-P$ phase transition, also
using the mean-field scale relations in the A-SWN regime $(0<p\le1)$.
In both figures, we have the comparison with the mean-field critical
exponents (MF) by the dotted lines, $\gamma=1.0$ and $\beta=\nu=0.5$.
\label{fig:9}}}
\end{figure}
\par\end{center}

The critical exponents are not independent one each other, but related
by simple scaling laws, as is the case with the hyperscaling law $d_{\textrm{eff}}=2\beta/\nu+\gamma/\nu$,
in which we have as a result the effective dimension $d_{\textrm{eff}}$
of the system. With this law, we see that the system has approximately
the same critical exponents in the A-SWN regime, as we are returned
that $d_{\textrm{eff}}\cong4.0$ with the mean field critical exponents,
and $d_{\textrm{eff}}\cong2.0$ following the data in the tables for
$p=0$, obtained with the scale relationships of Eqs. (\ref{eq:11}),
(\ref{eq:12}) and (\ref{eq:14}). 

The universality class can be defined by the set of exponents in the
phase transition, as in the case of the second-order phase transitions,
in which systems very different from each other can share the same
set of critical exponents. In general, these systems share the same
spatial dimension, symmetries, and range of interactions. Here, following
the set of critical exponents obtained at $p=0$, we have the same
universality class of the equilibrium Ising model in the regular square
lattice. However, in the A-SWN regime $(0<p\leq1)$, we have long-range
interactions in the system, and, due to its consequent set of critical
exponents, the system belongs mean-field universality class. By comparing
with the results obtained for the Ising model in the two-dimensional
A-SWN at the thermodynamic equilibrium regime \citep{24,key-1}, we
see that both the non-equilibrium model and the equilibrium model
have the same universality class, mean-field universality class, in
stationary critical behavior.

\subsection{Conclusions\label{subsec:Conclusions}}

In this work, we have developed MC simulations to study the thermodynamic
quantities and the critical behavior of the non-equilibrium Ising
model on a 2D A-SWN. By using the one- and two- spin flip competing
dynamics we reach the stationary state of the system at the non-equilibrium
regime. We have found two types of phase transitions, from $P$ to
$AF$ and from $P$ to $F$ phases, when the two-spin flip dynamic
prevails in the system, and when the one-spin flip prevails in the
system, respectively. To found the phases we have used the total and
staggered magnetizations per spin, and its respective susceptibility
and reduced fourth-order Binder cumulant, both as a function of the
competition parameter $q$. With the last two quantities are obtained
the critical points of the system and we built the phase diagrams
of the system. We have observed that increasing the coordination number
of the network by adding long-range interactions to our A-SWN, with
addition probability $p$, we also increase the regions of the ordered
phases on the diagram. Through the FSS arguments, we calculated the
critical exponents $\beta$, $\gamma$, and $\nu$, of the system,
and for the A-SWN regime $(0<p\leq1)$ we obtained the same exponents
of a system with mean-field critical behavior, except for $p=0$ as
expected, we obtained the critical exponents of the Ising model in
a regular square lattice. Thus, in the A-SWN regime, we have concluded
that the non-equilibrium system is in the mean-field universality
class, as the equilibrium system in the A-SWN \citep{24,key-1}. That
equivalence between the critical behavior of the equilibrium and non-equilibrium
models were already predicted and observed in other systems \citep{4,5,6,7,8}.
It is also important to specify that our results concerning the regions
of the phase diagram based on the changes in the critical points,
and the mean-field behavior is in agreement with the observed behavior
of disorder with shortcuts added to the Ising model in an SWN \citep{12,20,18,21,22,24,key-1}.

\begin{acknowledgments}
This work was partially supported by the Brazilian Agencies CNPq,
UFMT and FAPEMAT
\end{acknowledgments}

\end{document}